\documentclass[aps,prd,amssymb,eqsecnum,nofootinbib,floatfix, a4paper,twocolumn]{revtex4}

\usepackage{mathrsfs}
\usepackage{latexsym,amsmath,amsfonts,amssymb}
\usepackage{bm}

\usepackage{graphicx}

\usepackage{xcolor}  

\allowdisplaybreaks

\newcommand{\be}{\begin{equation}}
\newcommand{\ee}{\end{equation}}
\newcommand{\bea}{\begin{eqnarray}}
\newcommand{\eea}{\end{eqnarray}}

\newcommand{\sch}{{Schwarzschild \,}}
\newcommand{\cY}{{\cal Y}}

\def\al{\alpha}

\def\k{\kappa}
\def\hk{\hat \kappa}
\def\lam{{\lambda}}

\def\eps{\epsilon}
\def\d{\partial}
\def\l{\left(}
\def\r{\right)}

\def\t0{\tilde{0}}

\def\c_F{{\cal F}}
\def\c_R{{\cal R}}

\newcommand{\bg}{\begin{gather}}
\newcommand{\eg}{\end{gather}}
\newcommand{\bseq}{\begin{subequations}}
\newcommand{\eseq}{\end{subequations}}

\newcommand{\hi}{\hat{i}}
\newcommand{\hj}{\hat{j}}
\newcommand{\hak}{\hat{k}}
\newcommand{\htheta}{\hat{\theta}}
\newcommand{\hr}{\hat{r}}
\newcommand{\htt}{\hat{t}}

\begin{document}

\title{Spherically symmetric perturbations of a Schwarzschild black hole in torsion bigravity 
}

\author{Vasilisa \surname{Nikiforova}}
 
\affiliation{Institut des Hautes Etudes Scientifiques, 
91440 Bures-sur-Yvette, France}

\date{\today}

\begin{abstract}
Time-dependent spherically-symmetric perturbations of Schwarzschild black holes are studied within torsion bigravity, i.e., within generalized Einstein-Cartan theories where the dynamical torsion carries massive spin-2 excitation. We reduce linearized perturbations to a Zerilli-like equation. The structure of the potential entering the latter Zerilli-like equation has two important consequences. First, in order to avoid the presence of singularities in generic perturbations, one must restrict the range (or inverse mass) of the spin-2 excitation to be (essentially) smaller than the radius of the considered black hole.  Second, we then show that the Schwarzschild black hole is linearly stable against spherically-symmetric perturbations. 
\end{abstract}

\maketitle

\section{Introduction} \label{sec1}
The standard model of relativistic gravity, namely, General Relativity (GR), has been found to be in agreement with all experimental and observational data, on a very wide range of scales from a micron to cosmological scales. It is, however, important to be able to contrast GR predictions to predictions coming from alternative theories of gravity. Among alternative theories of gravity, two of them are distinguished by having, as spectrum, a {\it massive spin-2} excitation in addition to the Einstein-like {\it massless spin-2} one.  The first such theory is (ghost-free) {\it bimetric gravity} \cite{Hassan:2011zd}, which features two coupled space-time metrics, $g_{\mu\nu}$ and $f_{\mu\nu}$. The second one is {\it torsion bigravity} \cite{Damour:2019oru}, which is a generalized version of the Einstein-Cartan theory \cite{Cartan:1923zea, Cartan:1924yea, Cartan1925} comprising both a dynamical space-time metric $g_{\mu\nu}$, and a dynamical torsion ${T^\lam}_{\mu\nu}$.  General classes of dynamical torsion theories have been introduced at the end of the 1970s \cite{Sezgin:1979zf, Sezgin:1981xs, Hayashi:1979wj, Hayashi:1980av, Hayashi:1980ir, Hayashi:1980qp} and revived, within a cosmological context, in Refs.\cite{Nair:2008yh, Nikiforova:2009qr, Deffayet:2011uk, Nikiforova:2016ngy, Nikiforova:2017saf, Nikiforova:2017xww, Nikiforova:2018pdk}. 

The study of the physical properties of torsion bigravity has been recently initiated \cite{Damour:2019oru, Nikiforova:2020fbz, Nikiforova:2020sac, Nikiforova:2020oyp}.  
Static star-like torsion-hairy solutions were constructed \cite{Damour:2019oru}. Furthermore, it was found \cite{Nikiforova:2020fbz} that, contrary to the case of bimetric gravity, in torsion bigravity, there exists a smooth infinite-range limit which allows to consider small masses $\k$ of the spin-2 excitation without appealing to any Vainshtein-type mechanism. Then, following a path initiated in ghost-free bimetric gravity \cite{Volkov:2012wp, Babichev:2013una, Brito:2013xaa, Brito:2013wya},
black hole solutions were investigated in Refs.~\cite{Nikiforova:2020sac, Nikiforova:2020oyp}. {\it Torsionless} Einstein black-hole space-times are exact solution of torsion bigravity \cite{Nair:2008yh, Nikiforova:2009qr}. A no-hair theorem for {\it time-independent} linearized perturbations of a \sch black hole was proven, and non-asymptotically flat torsion-hairy black holes were constructed \cite{Nikiforova:2020sac}. In addition, it was shown that, in the infinite range limit, torsion bigravity admits torsion-hairy asymptotically flat black hole solutions \cite{Nikiforova:2020oyp}.    

The discovery  (in the infinite range limit) of the latter torsion-hairy asymptotically flat black holes poses the question whether they can be realized in Nature. In order to answer this question, we must investigate the stability of black hole solutions within torsion bigravity. In bimetric gravity,  it was found  \cite{Babichev:2013una, Brito:2013wya} that, when the mass of the spin-2 fluctuation\footnote{In this paper, the mass of the spin-2 excitation, i.e., the inverse of its range, will be denoted by $\k$.} is small enough, $\k r_h < 0.86$, where $r_h$ denotes the radius of the horizon, the \sch solution was unstable. The unstable mode was found to be spherically symmetric (being related to the Gregory-Laflamme instability \cite{Gregory:1993vy}). This leads us, in the present paper, to study the dynamical stability of \sch black hole against spherically symmetric perturbations within torsion bigravity.
We leave the study of non-spherically-symmetric perturbations to future work.

\section{Reminder of torsion bigravity formalism}

The fundamental fields of torsion bigravity are a space-time metric $g_{\mu\nu}$ (with mostly plus signature) and a metric-compatible ($\nabla ^{(A)}g=0$) affine connection ${A^\lam}_{\mu\nu}$ with torsion ${T^\lam}_{[\mu\nu]}$. The Lagrangian density of torsion bigravity reads
\bea \label{lag0}
L&=& c_R R[g]+  c_F F[g, A] \nonumber \\
&&+ c_{F^2}\left( F_{(\mu \nu)}[A] F^{(\mu \nu)}[A] - \frac13 F^2[g, A] \right) \nonumber \\
&& + c_{34}F_{[\mu \nu]}[A]F^{[\mu \nu]}[A] \,.
\eea
Here, $R[g]$ denotes the scalar curvature of $g_{\mu\nu}$, $F_{\mu\nu}[A] \equiv {F^\lam}_{\mu\lam\nu}[A]$ denotes the Ricci tensor of the connection ${A^\lam}_{\mu\nu}$, while $F[g, A] \equiv g^{\mu\nu}F_{\mu\nu}[A]$ denotes the corresponding Ricci scalar. The coupling constants $c_R$, $c_F$ and $c_{F^2}$ can be written as
\be
c_R = \frac{\lam}{1+\eta} \,, \quad  c_F = \frac{ \eta \lam}{1+\eta} \,, \quad c_{F^2} = \frac{ \eta \lam}{\k^2} \,,
\ee
where $\lam =c_F+c_R = \frac{1}{16\pi G_0}$ measures the gravitational coupling of the massless spin-2 fluctuation; $\eta = c_F/c_R$ is the ratio between the couplings of the massive and the massless spin-2 fluctuations, and $\k$ is the mass of the massive spin-2 fluctuation. [The massive spin-2 fluctuation is contained within the dynamical torsion ${T^\lam}_{[\mu\nu]}$.] The coupling constant $c_{34}$ multiplying the last contribution to the Lagrangian density \eqref{lag0} will enter intermediate equations of our analysis, but will drop out of our final results.  

As in previous works on dynamical torsion \cite{Sezgin:1979zf, Sezgin:1981xs, Hayashi:1979wj, Hayashi:1980av, Hayashi:1980ir, Hayashi:1980qp, Nair:2008yh, Nikiforova:2009qr, Deffayet:2011uk, Nikiforova:2016ngy, Nikiforova:2017saf, Nikiforova:2017xww, Nikiforova:2018pdk, Nikiforova:2020fbz, Nikiforova:2020sac, Nikiforova:2020oyp}, we introduce a vierbein ${e_{\hat{i}}}^\mu$, where the hatted latin indices $\hat{i}, \hat{j}, \ldots=0,1,2,3$ denote frame indices. In the following, we will use as basic field variables, the co-frame components ${e^{\hat{i}}}_\mu$, and the frame components ${A^{\hat{i}}}_{\hat{j}\hat{k}}$ of the connection. Let us also recall that the frame components ${T_{\hat{i}}}_{\hat{j}\hat{k}}=\eta_{\hi\hat{s}}{T^{\hat{s}}}_{\hat{j}\hat{k}}$ of the torsion tensor are related to the frame components ${K_{\hat{i}}}_{\hat{j}\hat{k}}=\eta_{\hi\hat{s}}{K^{\hat{s}}}_{\hat{j}\hat{k}}$ of the contorsion tensor, ${K^\lam}_{\mu\nu} \equiv {A^\lam}_{\mu\nu} - {\Gamma^\lam}_{\mu\nu}[g]$, via $T_{\hi [\hj\hak]} =  K_{\hi\hj\hak}- K_{\hi\hak\hj}$.

The explicit form of the field equations of torsion bigravity in terms of these variables can be found in \cite{Nikiforova:2018pdk} (see Eqs. (3.2) and (3.7) there).


\section{Perturbations of black holes in torsion bigravity} \label{sec2}
Vacuum (Ricci-flat) solutions of Einstein's equations are exact solutions of the field equations of torsion bigravity \cite{Nikiforova:2009qr}. In particular, stationary Einsteinian black hole solutions (\sch and Kerr) are torsionless solutions of torsion bigravity. Here we shall consider linearized perturbations of the \sch solution. The perturbations of \sch black holes in torsion bigravity are described by two tensor fields, the perturbation of the metric, $h_{\mu\nu}$, and the perturbation of the frame components of the connection, ${a^{\hat{i}}}_{\hat{j}\hat{k}}$:
\bea \label{linearization}
g_{\mu\nu}(t,r,\theta,\phi) &=& g_{\mu\nu}^S(r,\theta,\phi)  +\varepsilon h_{\mu\nu}(t,r,\theta,\phi) + O(\varepsilon^2) \,,  \nonumber \\
 {A^{\hat{i}}}_{\hat{j}\hat{k}}(t,r,\theta,\phi) &=&  {A^{\hat{i}\,S}}_{\hat{j}\hat{k}}(r,\theta,\phi) + \varepsilon {a^{\hat{i}}}_{\hat{j}\hat{k}}(t,r,\theta,\phi) + O(\varepsilon^2) \,. \nonumber \\
\eea
Here the superscript $S$ denotes \sch background values. We can decompose the perturbations both in frequency space and in tensorial harmonics:
\bea
h_{\mu\nu}(t,r,\theta,\phi) &=& \sum_{l,m} \int_{-\infty}^{\infty} d\omega \, e^{-i\omega t} [h_{\mu\nu}^{{\rm even},\,lm}(\omega,r,\theta,\phi) \nonumber \\
&& +h_{\mu\nu}^{{\rm odd},\,lm}(\omega,r,\theta,\phi) ] \,,
\eea
\bea
{a^{\hat{i}}}_{\hat{j}\hat{k}}(t,r,\theta,\phi) &=& \sum_{l,m} \int_{-\infty}^{\infty} d\omega \, e^{-i\omega t} [{a^{\hat{i}}}_{\hat{j}\hat{k}}^{\,{\rm even},\,lm}(\omega,r,\theta,\phi) \nonumber \\
&& +{a^{\hat{i}}}_{\hat{j}\hat{k}}^{\,{\rm odd},\,lm}(\omega,r,\theta,\phi) ] \,.
\eea
Here, we decomposed the perturbations in even-parity ones and odd-parity ones. For the reasons explained in the Introduction, we shall only consider here spherically-symmetric perturbations: $(l,\,m)=(0,\,0)$.

Exact time-dependent spherically-symmetric solutions of torsion bigravity are described by {\it ten} variables. First, there are two metric variables, $\Phi(t,r)$ and $\Lambda(t,r)$, such that
\be \label{ds2}
ds^2=-e^{2\Phi}dt^2 + e^{2\Lambda}dr^2 + r^2\left( d\theta^2+\sin^2\theta\, d\phi^2 \right) \,.
\ee
As is always possible  \cite{Choquet-Bruhat:2014okh} for generic time-dependent spherically-symmetric metrics, we used here  a Schwarzschild-type coordinate system with $g_{tr}=0$ and $g_{\theta\theta}=g_{\phi\phi}/\sin^2\theta=r^2$. The connection components will then refer to the orthonormal (co-)frame 
$\theta^{\hat{i}}= {e^{\hat{i}}}_\mu dx^\mu$ ($\hat{i}=\hat{0},\hat{1},\hat{2},\hat{3}$, or $\hat{t}, \hat{r}, \hat{\theta}, \hat{\phi}$) with
\be \label{frame}
\theta^{\hat{0}}=e^{\Phi}dt \;, \quad \theta^{\hat{1}}=e^{\Lambda}dr \;, \quad \theta^{\hat{2}}=r d\theta \;, \quad \theta^{\hat{3}}= r \sin\theta d\phi \;.
\ee
Besides the two metric variables $\Phi(t,r)$ and $\Lambda(t,r)$, there are four {\it even-parity} connection variables, $V(t,r)$, $W(t,r)$, $X(t,r)$ and $Y(t,r)$,
\bea \label{VWXY}
V(t,r) & \equiv &{A^{\hat{1}}}_{\hat{0}\hat{0}}= e^{-\Lambda}\Phi^{\prime} + {T^{\hat{0}}}_{\hat{1}\hat{0}}  \,,  \nonumber \\
W(t,r) & \equiv &{A^{\hat{1}}}_{\hat{2}\hat{2}}= - \frac{e^{-\Lambda}}{r} - {T^{\hat{2}}}_{\hat{1}\hat{2}}  \,,  \nonumber \\
X(t,r) & \equiv &{A^{\hat{1}}}_{\hat{0}\hat{1}}= e^{-\Phi}\d_t \Lambda + {T^{\hat{1}}}_{\hat{0}\hat{1}}   \,,  \nonumber \\
Y(t,r) & \equiv &{A^{\hat{2}}}_{\hat{0}\hat{2}}= {T^{\hat{2}}}_{\hat{0}\hat{2}}   \,,  \nonumber \\
\eea
and four {\it odd-parity} connection variables, $C_1(t,r)$, $C_2(t,r)$, $C_3(t,r)$ and $C_4(t,r)$,
\bea \label{OddVar}
C_1(r,t) & \equiv & {A^{\hat{3}}}_{\hat{2}\hat{0}}=-{A^{\hat{2}}}_{\hat{3}\hat{0}} = {T^{\hat{2}}}_{\hat{0}\hat{3}} - \frac{1}{2} {T^{\hat{0}}}_{\hat{2}\hat{3}}  \,,  \nonumber \\
C_2(r,t) & \equiv & {A^{\hat{3}}}_{\hat{2}\hat{1}}=-{A^{\hat{2}}}_{\hat{3}\hat{1}} = {T^{\hat{2}}}_{\hat{1}\hat{3}} + \frac{1}{2} {T^{\hat{1}}}_{\hat{2}\hat{3}}  \,,  \nonumber \\
C_3(r,t) & \equiv & -{A^{\hat{0}}}_{\hat{2}\hat{3}}={A^{\hat{0}}}_{\hat{3}\hat{2}}=-{A^{\hat{2}}}_{\hat{0}\hat{3}}={A^{\hat{3}}}_{\hat{0}\hat{2}} \nonumber \\
&&=- \frac{1}{2} {T^{\hat{0}}}_{\hat{2}\hat{3}}  \,,  \nonumber \\
C_4(r,t) & \equiv & -{A^{\hat{1}}}_{\hat{2}\hat{3}}={A^{\hat{1}}}_{\hat{3}\hat{2}}={A^{\hat{2}}}_{\hat{1}\hat{3}}=-{A^{\hat{3}}}_{\hat{1}\hat{2}} \nonumber \\
&&=- \frac{1}{2} {T^{\hat{1}}}_{\hat{2}\hat{3}}  \,.  \nonumber \\
\eea
In Eqs. \eqref{VWXY} and \eqref{OddVar}, ${T^{\hat{i}}}_{\hat{j}\hat{k}}=-{T^{\hat{i}}}_{\hat{k}\hat{j}}$ denote the frame components of the torsion tensor.
The definitions of the four connection variables $C_1, \ldots, C_4$ all depend on the choice of an orientation within the 2-sphere $(\hat{2}, \hat{3})=(\hat{\theta}, \hat{\phi} )$, hence their odd-parity character.

The background (i.e., Schwarzschild) values of the odd-parity variables all vanish: $0=C^S_1 =C^S_2=C^S_3=C^S_4$.
The odd-parity spherically-symmetric perturbations are studied in Appendix B and shown there to be trivial. In the following, we focus on even-parity perturbations. 

The unperturbed components of the metric and (even-parity) connection variables describing a \sch black hole are
\bea \label{schw}
\Phi_S(r) &=&  +\frac12 \ln \left(1 - \frac{r_h}{r}\right)\,, \nonumber\\
\Lambda_S(r) &=& -\frac12 \ln \left(1 - \frac{r_h}{r}\right) \,, \nonumber\\
V_S(r)  &=& \frac12 \frac{r_h}{r^2}\left( 1-\frac{r_h}{r} \right)^{-1/2}\,, \nonumber\\
W_S(r) &=&  -\frac{1}{r} \left( 1-\frac{r_h}{r} \right)^{1/2}  \,, \nonumber \\
X_S(r) &=&0  \,, \nonumber \\
Y_S(r) &=&0  \,,
\eea
where $r_h$ denotes the \sch radius. 

We will use the following specific notation for the frequency-space linearized \sch perturbations,
\be \label{o-list}
\phi_o (\omega, r), \Lambda_o (\omega, r),  V_o (\omega, r), W_o (\omega, r), X_o (\omega, r), Y_o (\omega, r)\,, 
\ee
where, for instance,
\be \label{o-def}
\Phi (t, r) = \Phi_S (r) + \varepsilon \int_{-\infty}^{\infty} d\omega \, e^{-i\omega t} \phi_o (\omega, r) + O(\varepsilon^2) \,.
\ee

In what follows, we use a $\prime \equiv \d_r$ to denote the $r$-derivative, and $\dot{} \equiv \d_t$ to denote the derivative with respect to $t$.

\section{Reduction of the linearized field equations to a system of two first-order radial equations} \label{sec3}

There are fourteen 
exact field equations describing time-dependent spherically-symmetric torsion bigravity configurations ($\Phi(t,r)$, $\Lambda(t,r)$, $V(t,r), \ldots$). They have been written down by Rauch and Nieh \cite{Rauch:1981tva}. See (4.3a)--(4.3f), (4.5a)--(4.5d) and (4.6a)--(4.6d) there. [Beware that, contrary to the latter equations, the rewritten Eqs.~(4.4a)--(4.4f) and (6.2a)--(6.2e) contain some misprints. See Appendix \ref{apA} for details.] Nine of these exact field equations contain the odd-parity variables $C_1, \ldots, C_4$ only quadratically, while five of them are linear in the odd-parity variables. The former nine even-parity equations are given in Appendix \ref{apA} (neglecting to write contributions quadratic in $C_1, \ldots, C_4$ which do not enter the linearized level).  The corresponding nine (even-parity) linearized perturbed equations (using Eq.\eqref{o-def}) for the frequency-space variables \eqref{o-list} can be found in the {\it Supplemental Material}. 

Among the linearized even-parity equations, several of them contain second-order radial derivatives of the field variables. Namely, the linearized version of Eq.~\eqref{EG3LinOsc1} contains $\phi_o^{\prime\prime}(r)$, the linearized Eq.~\eqref{ET1LinOsc1} and Eq.~\eqref{ET4LinOsc1}  contain $\phi_o^{\prime\prime}(r)$, $V_o^{\prime\prime}(r)$ and $W_o^{\prime\prime}(r)$, and the linearized Eq.~\eqref{ET3LinOsc1} contains $Y_o^{\prime\prime}(r)$. Actually, $V_o^{\prime\prime}(r)$ and $W_o^{\prime\prime}(r)$ always appear in the single combination $ V_{o}^{\prime\prime}+W_{o}^{\prime\prime}$. In addition, $\phi_o$ never appears undifferentiated. [This is linked to the residual gauge invariance $\Phi(t,r)\to \Phi(t,r)+f(t)$ of the \sch coordinate gauge used in Eq.\eqref{ds2}]. As a consequence of these properties, one can transform the set of nine linearized field equations Eq.~\eqref{EG1LinOsc1}-Eq.~\eqref{ET4LinOsc1} into an equivalent set of eleven {\it first-order} differential equations by introducing the three auxiliary variables 
\bea  \label{auxvar}
F_o(\omega,r) &\equiv & \phi_o^{\prime}(\omega,r) \,, \nonumber \\
 Z_{o}(\omega,r) &\equiv & Y_{o}^{\prime}(\omega,r) \,,  \nonumber \\
 p_{o}(\omega,r) &\equiv & V_{o}^{\prime}(\omega,r)+W_{o}^{\prime}(\omega,r) \,.
\eea
More precisely, one finds that the {\it eight} variables
\be  \label{8var}
\left[{\cal Y}_{i}(\omega,r)\right]_{i=1,\ldots,8} \equiv 
 \{F_o,\, \Lambda_o, \, V_o,\, W_o, \, X_o, \, Y_o, \, Z_{o}, \, p_{o}  \} \,,
 \ee
must satisfy a set of {\it eleven} first-order linear differential equations with respect to $r$, of the form
\be
 A_{\alpha i}(\omega,r)\, {\cal Y}_i^{\prime}(\omega,r) + B_{\alpha i}(\omega,r)\, {\cY}_i(\omega,r) =0 \label{mat11eq0} \,.
 \ee
Here the index $\alpha=1,\ldots, 11$ labels the eleven linearized field equations, while the index $i=1,\ldots,8$ labels the eight frequency-space perturbed field variables \eqref{8var}. We use Einstein's summation convention on all repeated indices (here: $i=1,\ldots,8$). Two of these equations are consequences of Eqs.~\eqref{auxvar}, namely,
 \bea
&& Y_{o}^{\prime}(\omega,r)-Z_{o}(\omega,r) = 0  \,,  \nonumber \\
 && V_{o}^{\prime}(\omega,r)+W_{o}^{\prime}(\omega,r)-p_{o}(\omega,r) =0  \,.
 \eea
Let us display here, for concreteness, another equation in the system \eqref{mat11eq0}:
\begin{widetext}

\bea
&& - \frac{
 \sqrt{r - r_h}
   r_h \eta }{
 \k^2 r^{7/2}}\left[V_o^{\prime}(\omega,r) + W_o^{\prime}(\omega,r)\right] +  \frac{4 \k^2 r^3 (r - r_h) - r_h^2 \eta (1 + \eta) }{
 2 \k^2 r^5 (1 + \eta)}F_o(\omega,r)  \nonumber \\
  && +\eta \frac{ 4 \k^2 r^3 (r - r_h) - (2 r - 
       r_h) (1 + \eta)r_h }{
 2 \k^2 r^{9/2} \sqrt{
  r - r_h} (1 + \eta)}V_o(\omega,r) - \eta\frac{ 2 \k^2 r^3 (2 r - r_h) + (6 r - 
       7 r_h) (1 + \eta)r_h }{
 2 \k^2 r^{9/2} \sqrt{r - r_h} (1 + \eta)}W_o(\omega,r)  \nonumber \\
  && - \frac{
 i \omega r_h \eta  X_o(\omega,r)}{\k^2 r^{5/2} \sqrt{r - r_h}} - \frac{
 i \omega\eta (-2 \k^2 r^3 + r_h + r_h \eta)  }{
 \k^2 r^{5/2} \sqrt{
  r - r_h} (1 + \eta)}Y_o(\omega,r) - \frac{4 \k^2 r^4 + 
    3 r_h^2 \eta (1 + \eta) }{
 2 \k^2 r^6 (1 + \eta)}\Lambda_o(\omega,r)  = 0 \,.
\eea
\end{widetext}

The coefficients $A_{\alpha i}(\omega,r)$ entering Eqs.~\eqref{mat11eq0} are, generally speaking, first-order polynomials in $\omega$, while the coefficients $B_{\alpha i}(\omega,r)$ are second-order polynomials in $\omega$. This property comes from the fact that the original field equations were second-order in time derivatives. 

In the following, we will think of the system \eqref{mat11eq0} in matrix form. Namely, 
\be \label{mat11eq}
 A\, {\cal Y}^{\prime} + B\, {\cY} =0  \,,
 \ee
where $A$ and $B$ are $11\times 8$ matrices, and $\cY$ is an 8-dimensional column vector. 

This radial evolution system implies a certain number of algebraic constraints on the variables $\cY$. First, one
obtains {\it primary constraints} (in the sense of Dirac). These constraints are linked to the rank of the matrix $A$. We find that the rank of $A$ is six. This implies, in particular, that the {\it left} null-space of the $11\times 8$ matrix $A$ is five-dimensional.  Indeed, any left null eigenvector $v_{\al}$ of $A$, namely, any solution of the equation
\be
v_{\al} \, A_{\al i} =0 \,,
\ee
implies (by contracting $v_{\al}$ with the field equations \eqref{mat11eq0}) the corresponding algebraic constraint 
\be  \label{Cprimary}
C^{{\rm primary}}(v) \equiv v_{\al} \, B_{\al i}{\cal Y}_i =0 \,.
\ee
There are five such  {\it primary}  constraints corresponding to the five-dimensional nature of the left null-space of the matrix $A$, or equivalently, to the right null-space of the {\it transpose} matrix $A^T$. Explicit computation of the right null-space of $A^T$ shows that the five corresponding primary constraints $C^{{\rm primary}}(v)$, Eq.~\eqref{Cprimary}, are independent. Indeed, we find that the five constraints $C^{{\rm primary}}(v)$ on the eight variables $\cY$ can be solved for $F_o,\,\Lambda_o,\,Y_o,\,Z_o, \,p_o$ in terms of the three residual variables, 
\be \label{3var}
({\cal Z}_{a})_{a=1,2,3} \equiv \{ V_o, \, W_o, \, X_o \} \,.
\ee

After substituting the solutions 
\be \label{sol}
F_o ({\cal Z}_{a}), \, \Lambda_o({\cal Z}_{a}),\,Y_o({\cal Z}_{a}),\,Z_o({\cal Z}_{a}),\,p_o({\cal Z}_{a})
\ee
 of the five primary constraints in the eleven original equations \eqref{mat11eq0}, we obtain a system of the form
\be \label{secondarysys}
C_{\al a} \,{\cal Z}_{a}^\prime + D_{\al a} \,{\cal Z}_{a} =0 \,.
\ee
Here the index $\al=1,\ldots,11$ takes eleven values, while $a=1,2,3$.

Following the Dirac approach, we must now study the rank of the matrix $C$ appearing in \eqref{secondarysys}
to know how many equations are independent, and how many secondary constraints they imply. By explicit computation one finds that the rank of the $11\times 3$ matrix $C$ is equal to two. This means that the {\it left} null-space of $C$ is nine-dimensional. Denoting by $w$ any left null-eigenvector of the matrix $C$, we thereby get nine {\it secondary} constraints,
\be  \label{Csecondary}
C^{{\rm secondary}}(w) \equiv w_{\al} \, D_{\al a}{\cal Z}_a =0 \,,
\ee
on the three variables $({\cal Z}_{a})=\{ V_o, \, W_o, \, X_o \}$. Explicit computation shows that these nine constraints are proportional to each other. Thus, there is only one independent {\it secondary} constraint among the three variables, $V_o, \, W_o, \, X_o$. We then solve this single secondary constraint for $X_o$, say,
\be \label{Xsol}
X_o=X_o^{\rm sol}(V_o,\,W_o) \,.
\ee
Inserting this relation in the previous solutions \eqref{sol} yields six solutions:
\bea \label{AC6}
&& F_o=F_o^{\rm sol} (V_o, W_o), \, \Lambda_o=\Lambda_o^{\rm sol} (V_o, W_o),\nonumber \\
&& Y_o=Y_o^{\rm sol} (V_o, W_o), \,
 Z_o=Z_o^{\rm sol} (V_o, W_o),\nonumber \\
 && p_o=p_o^{\rm sol} (V_o, W_o), \, X_o=X_o^{\rm sol} (V_o, W_o) \,.
\eea
Substituting this solution in the original set of equations, one finds that the full set of perturbed equations
is equivalent to a system of {\it two} first-order differential equations for the two variables, $V_o$ and $W_o$. Say,
\bea \label{2sys}
V_o^{\prime}(\omega,r) &=& C_{VV}(\omega,r) V_o(\omega,r) + C_{VW}(\omega,r) W_o(\omega,r) \,, \nonumber \\
W_o^{\prime}(\omega,r) &=& C_{WV}(\omega,r) V_o(\omega,r) + C_{WW}(\omega,r) W_o(\omega,r) \,. \nonumber  \\
\eea
The coefficients $C_{VV}(\omega,r), C_{VW}(\omega,r),\ldots$ entering Eqs.~\eqref{2sys} are rational functions of $\omega^2$. More precisely, $C_{VV}$, $C_{WV}$ and $C_{WW}$ are of the form 
\be
\frac{a_0(r) +a_2(r)\omega^2}{b_0(r)+b_2(r)\omega^2} \,,
\ee
while $C_{VW}$ is of the form
\be
C_{VW}=\frac{a_0(r) +a_2(r)\omega^2+a_4(r)\omega^4}{b_0(r)+b_2(r)\omega^2} \,.
\ee
The various coefficients $a_n(r)$, $b_n(r)$ are algebraic functions of $r$ (involving $\sqrt{r-r_h}$) and polynomials in $\kappa^2$ and $\eta$. The explicit expressions of these coefficients are given in the Supplemental Material. 

The reduction of the full set of perturbed equations to a system of two first-order differential equations 
(whose general solution is parametrized by two initial data) does correspond to the expected degrees
of freedom for time-dependent spherically symmetric solutions of the field content of torsion
bigravity, namely a massless spin-2 and a massive spin-2 one. Indeed, by Birkhoff's theorem, the massless 
spin-2 time-dependent spherically symmetric sector is trivial, and it is easily seen that time-dependent spherically-symmetric massive spin-2 excitations must involve two initial data. 
 Let us note in passing that this is a further confirmation of the absence of Boulware-Deser sixth degree of freedom \cite{Boulware:1973my}. A complementary side of this result is that the odd-parity time-dependent spherically-symmetric perturbed sector is expected to be trivial. This is indeed explicitly checked in Appendix B.

\section{Reduction to a Zerilli-like equation} \label{sec4}
The non-polynomial dependence on $\omega$ of the frequency-domain system \eqref{2sys}  does not allow one to easily analyze the behavior of the perturbations if we wanted to analyze them in the time domain. The question then arises whether it is possible to transform our system \eqref{2sys} into a Zerilli-type equation, i.e., an equation of the form 
\be \label{EqPhi}
 \frac{\d^2}{\d r_*^2} \varphi_{\omega}(r_*)= \left( V[r(r_*)]-\omega^2 \right)\, \varphi_{\omega}(r_*)\,
\ee
with a frequency-independent potential $V[r(r_*)]$. Here, as usual, $r_*$ denotes the tortoise radial coordinate,
\be
r_*=r+r_h \ln{\left(r/r_h-1\right)} \,, \quad \frac{dr_*}{dr}=\frac{r_h}{r-r_h}\,.
\ee 
Let us recall indeed that, after transforming to the time domain, namely, 
\be
\varphi(t,r_*)=\int_{-\infty}^{+\infty} d\omega e^{-i \omega t} \varphi_{\omega}(r_*) \,,
\ee
Eq.~\eqref{EqPhi} reads 
\be \label{EqPhit}
 \frac{\d^2}{\d r_*^2} \varphi(t,r_*) - \frac{\d^2}{\d t^2} \varphi(t,r_*)= V[r(r_*)] \, \varphi(t,r_*)\,. 
\ee
The latter equation exhibits the fact that the (front) velocity of the black hole perturbations is equal to the speed of light. 

The transformation from the system \eqref{2sys} to an equation of the type \eqref{EqPhi} comprises two steps. To motivate the first step, let us recall  a result of Refs.~\cite{Nikiforova:2009qr, Deffayet:2011uk} concerning 
perturbations of Einstein spaces in torsion gravity. If one denotes the following (symmetrized) combination 
of the frame components $F_{\hi\hj}$ of the Ricci tensor  of the connection ${A^{\hi}}_{\hj\mu}$ as
\be
U_{\hi\hj} \equiv F_{(\hi\hj)} - \frac{1}{6}F \, \eta_{\hi\hj}
\ee
its perturbed value around Einstein spaces, namely,
\be \label{udef}
u_{\hi\hj} \equiv U_{\hi\hj}^{(1)}  \equiv F_{(\hi\hj)}^{(1)} - \frac{1}{6}F^{(1)}\eta_{\hi\hj} \,,
\ee
satisfies a generalized Fierz-Pauli equation comprising both a mass term and an additional
coupling to the Weyl tensor of the background, namely
\be \label{Wcoupling}
\kappa^2 (u_{\hi\hj}- u\, \eta_{\hi\hj} ) + (1+\eta) W_{\hi \hk \hj \hat{l}} u^{\hk\hat{l}} \,. 
\ee
This result indicates that
it will be useful to replace the two basic (connection-related) variables $V_o$, $W_o$ entering the system
\eqref{2sys}  by two other variables more directly connected with the auxiliary Fierz-Pauli-like variables $u_{\hi\hj}$.
An analog approach has been used when considering perturbed black holes within  bimetric gravity \cite{Brito:2013wya}. The latter reference used the combination 
$h_{\mu \nu}^{(m)} \propto M_g \delta f_{\mu \nu} - C M_f \delta g_{\mu \nu}$ of the perturbations
of the two  metric tensors $g_{\mu \nu}$,  $f_{\mu \nu}$ that satisfies a Fierz-Pauli-like equation as a starting
point to construct a variable  $ \varphi(t,r_*)$ satisfying a Zerilli-like equation \eqref{EqPhi}.

Our first step will therefore be to derive the explicit expressions of the torsion-bigravity variables $u_{\hi\hj}$
in terms of our two basic variables $V_o$, $W_o$. In a generic time-dependent spherically-symmetric situation,
the non-zero components of $u_{\hi\hj}$ are $u_{\hat{0}\hat{0}}$,  $u_{(\hat{1}\hat{0})}$,  $u_{\hat{1}\hat{1}}$,  and $u_{\hat{2}\hat{2}}= u_{\hat{3}\hat{3}}$. Within the usual Regge-Wheeler-Zerilli \cite{Regge:1957td,Zerilli:1970se} setting, these frame components of
a generic metric perturbation are respectively denoted as $H_0$, $H_1$, $H_2$ and $K$.
Within our perturbed torsion-bigravity setting, the results obtained in the previous sections has shown
that any perturbed variable can be finally expressed (by using the algebraic constraints \eqref{AC6}, together
with the differential constraints \eqref{2sys})  as a linear combination of $V_o$ and $W_o$. 
This fact shows in particular that the four metric-like variables $u_{\hat{0}\hat{0}}$,  $u_{(\hat{1}\hat{0})}$,  $u_{\hat{1}\hat{1}}$,  and $u_{\hat{2}\hat{2}}$ (or equivalently  $H_0$, $H_1$, $H_2$ and $K$) satisfy
two algebraic constraints. It is therefore enough to  chose two independent components of $u_{\hi\hj}$ and to express them in terms of our basic (connection-related) variables $V_o$ and $W_o$. Inspired by the results of 
Zerilli \cite{Zerilli:1970se} (and their bimetric-gravity analogs \cite{Brito:2013wya}), we chose to work with
the two variables $u_{(\hat{1}\hat{0})} = H_1$ and $u_{(\hat{2}\hat{2})}=K$. These are the perturbed values of the exact
Ricci components $U_{\hi\hj} \equiv F_{(\hi\hj)} - \frac{1}{6}F\eta_{\hi\hj}$ for $\hi\hj= \hat{0}\hat{1}$ and $\hi\hj=\hat{2}\hat{2}$. An explicit
calculation yields
\bea \label{HKdef}
 &&U_{(\hat{1}\hat{0})}(t,r)=-\d_rY(t,r) e^{-\Lambda(t, r)} + \d_tW(t,r) e^{-\Phi(t,r)} \nonumber \\
 && - 
 W(t,r) X(t,r) + \left[-\frac{e^{-\Lambda(t,r)}}{r} + V(t,r)\right] Y(t,r) \,, \nonumber \\
 &&U_{(\hat{2}\hat{2})}=\frac{1}{3} \left[\d_rV(t,r) e^{-\Lambda(t,r)} + \d_rW(t,r) e^{-\Lambda(t,r)} \right. \nonumber \\
 && - 
   \d_tX(t,r) e^{-\Phi(t,r)} + \d_tY(t,r) e^{-\Phi(t,r)} + \frac{2}{r^2} \nonumber \\
   && + 
   F(t,r) e^{-\Lambda(t,r)} V(t,r) - 
   2 W(t,r)^2 + 2 Y(t,r)^2 \nonumber \\
   && + \left(\frac{e^{-\Lambda(t,r)}}{r} + V(t,r)\right) W(t,r)  \nonumber \\
   && \left. + 
   X(t,r) \left(-\d_t\Lambda e^{-\Phi(t,r)} + Y(t,r)\right)  \right] \,.
\eea 
Linearizing these exact expressions and passing to frequency-space yields the following primary expressions for
$u_{(\hat{1}\hat{0})}(\omega,r) \equiv H_1(\omega,r)$ and $u_{\hat{2}\hat{2}}(\omega,r) \equiv K(\omega,r)$ in terms of $F_o,\Lambda_o, V_o, W_o, X_o, Y_o$ and their radial derivatives:
\bea \label{HKdef}
&&H_1(\omega,r)=\left[-2 i r^2 \omega W_o(r) + 2 (r - r_h) X_o(r) - 2 r Y_o(r) \right. \nonumber \\
&& \left.+ 
 3 r_h Y_o(r) - 2 r(r-r_h) Y_o^{\prime}(r) \right]/\left(2 \sqrt{r^3 (r - r_h)}\right) \,, \nonumber \\
&&K(\omega,r)= \left[\sqrt{r (r - r_h)} r_h F_o(r) + r (  3 r_h-2 r) V_o(r)\right.  \nonumber  \\
&& + 10 r^2 W_o(r) -  9 r r_h W_o(r) + 2 i r^3 \omega X_o(r) \nonumber \\
&& - 2 i r^3 \omega Y_o(r) +  3 r_h \sqrt{1 - \frac{r_h}{r}} \Lambda_o(r)  + 
 2 r^2(r-r_h) V_o^{\prime}(r) \nonumber \\
 &&\left.  + 
 2 r^2(r-r_h) W_o^{\prime}(r) \textcolor{white}{\frac{i}{1}} \right] / 
 \l 6 \sqrt{r^5 (r - r_h)} \r \,.
\eea
Substituting the algebraic relations \eqref{AC6} of Sec.~\ref{sec3} into these expressions,
and using the differential conditions \eqref{2sys}, finally yields the following explicit algebraic expressions  for $H_1$ and $K$ in terms of $V_o$ and $W_o$:
\bea \label{HKvsVW}
&& H_1 = -\frac{2 i \omega}{\sqrt{1 - \frac{r_h}{r}} \, \eta \left[\k^2 r^3 - r_h (1 + \eta)\right] } \times \nonumber \\
&& \left\{-2 (r - r_h) r_h (1 + \eta) \left[\k^4 r^6 + r_h^2 \eta (1 + \eta) \right. \right.  \nonumber \\
     && \left. - 
     \k^2 r^3 r_h (1 + 2 \eta)\right] V_o(R) \nonumber \\
     && + \left[r_h^2 \eta (1 + \eta)^2 (2 r r_h - r_h^2 + 
        4 r^4 \omega^2) \right. \nonumber \\
        && - 
     2 \k^2 r^3 r_h (1 + \eta) (r_h^2 (1 - 4 \eta) + 
        r r_h ( 5 \eta-1) + 4 r^4 \eta \omega^2) \nonumber \\
        && + 
     \k^4 r^6 (-r_h^2 (4 + 7 \eta) + 4 r (r_h + 2 r_h \eta) \nonumber \\
     && \left. \left.+ 
        4 r^4 \eta \omega^2)\right] W_o(r)\right\} \nonumber \\
     &&  / \left[r_h (1 + \eta) (-10 r r_h + 9 r_h^2 - 4 r^4 \omega^2) \right. \nonumber \\
     && \left. + 
 \k^2 r^3 (4 r r_h - 3 r_h^2 + 4 r^4 \omega^2)\right] \,, \nonumber \\
&& K = \frac{2 \sqrt{r - 
   r_h}}{r^{3/2} \eta} \left\{2 (r - r_h) \left[\kappa^4 r^6 + r_h^2 \eta (1 + \eta) \right. \right. \nonumber \\
   && \left.  - 
       \k^2 r^3 r_h (1 + 2 \eta)\right] V_o(r) + [-4 \k^4 r^6 (r - r_h) \nonumber \\
       && + 
       r_h \eta (1 + \eta) (-2 r r_h + r_h^2 - 4 r^4 \omega^2) + 
       \k^2 r^3 r_h^2 (2 + \eta)  \nonumber \\
       && \left. \left. + 
       \k^2 r^3 (  4 r^4 \eta \omega^2 -2 r r_h)\right] W_o(r)\right\}/\{ r_h (1 + \eta) (9 r_h^2  \nonumber \\
       && -10 r r_h  - 
       4 r^4 \omega^2) + 
    \k^2 r^3 (4 r r_h - 3 r_h^2 + 4 r^4 \omega^2)\} \,.
\eea
Differentiating the latter expressions for $H_1$ and $K$ yields the values of $H_1^{\prime}$ and $K^{\prime}$ in terms of $V_o, W_o, V_o^{\prime}, W_o^{\prime}$. Then, using the differential system \eqref{2sys} and the inverse  relations $V_o=V_o(H_1,K)$, $W_o=W_o(H_1,K)$ obtained by solving the system \eqref{HKvsVW}, one obtains a linear system of two differential equations for $H_1$ and $K$.  Rewriting this system in terms of $K$ and $\tilde{H}=H_1\omega^{-1}$ leads to a system of the form
\bea \label{KHtsys}
K^{\prime}&=&C_{KK}(\omega,r)K+C_{KH}(\omega,r)\tilde{H} \,, \nonumber \\
\tilde{H}^{\prime}&=&C_{HK}(\omega,r)K+C_{HH}(\omega,r)\tilde{H} \,.
\eea
A crucial feature of the coefficients entering this differential system is that each of them is now found to have a simple {\it linear dependence} on $\omega^2$, namely
\be
C_{KK}(\omega,r)=a_{KK}(r)+\omega^2 b_{KK}(r), \; {\rm etc}\,.
\ee
The latter linear dependence on $\omega^2$ allows us to apply the procedure introduced by Zerilli \cite{Zerilli:1970se} in the General Relativity setting. This procedure consist in looking for a $2\times 2$ matrix $M(r)$ (depending only on $r$), say 
\be \label{tranm}
M(r)=\begin{pmatrix} f(r) & g(r) \\
 h(r) & k(r) \end{pmatrix} \, ,
\ee
such that the transformation 
\be \label{ZTransform}
\begin{pmatrix} K \\ \tilde{H} \end{pmatrix}=M(r) \begin{pmatrix}\varphi \\ \psi \end{pmatrix} 
\ee
maps the system \eqref{KHtsys} on a first-order system of the form
\be \label{SysPhi}
\begin{cases}
\displaystyle{{r-r_h}\over{r}}\varphi^{\prime}=\d_{r_*}\varphi=\psi \,, \nonumber \\
\displaystyle{{r-r_h}\over{r}}\psi^{\prime}=\d_{r_*}\psi= \left(V(r)-\omega^2\right)\varphi \,
\end{cases}
\ee
(this system is equivalent to Eq.~\eqref{EqPhit})
Writing the conditions following from this procedure we found an explicit solution given by the expressions 
\bea
&f(r)&= \frac{k(
    r)}{2r^4 D_f(r)} i (r - r_h) \l 4 \k^8 r^{13} + \k^{10} r^{15} + 2 \k^6 r^{10} r_h \right. \nonumber \\
&& - 6 \k^8 r^{12} r_h - 
   \k^6 r^9 r_h^2 + 4 \k^6 r^{10} r_h \eta - 3 \k^8 r^{12} r_h \eta \nonumber \\
   && - 
   18 \k^4 r^7 r_h^2 \eta - \k^6 r^9 r_h^2 \eta + 
   4 \k^2 r^4 r_h^3 \eta + 19 \k^4 r^6 r_h^3 \eta \nonumber \\
   && - 
   5 \k^2 r^3 r_h^4 \eta - 18 \k^4 r^7 r_h^2 \eta^2 + 
   2 \k^6 r^9 r_h^2 \eta^2 \nonumber \\
   && + 12 \k^2 r^4 r_h^3 \eta^2 + 
   21 \k^4 r^6 r_h^3 \eta^2 + 2 r r_h^4 \eta^2 - 
   17 \k^2 r^3 r_h^4 \eta^2 \nonumber \\
   && - r_h^5 \eta^2 + 
   8 \k^2 r^4 r_h^3 \eta^3 + 2 \k^4 r^6 r_h^3 \eta^3 + 
   4 r r_h^4 \eta^3 \nonumber \\
   && - 15 \k^2 r^3 r_h^4 \eta^3 - r_h^5 \eta^3 + 
   2 r r_h^4 \eta^4 - 3 \k^2 r^3 r_h^4 \eta^4 \nonumber \\
   && \left. + r_h^5 \eta^4 + 
   r_h^5 \eta^5 \r  \,, \\
   &g(r)& = \frac{i (r - r_h) k(r)}{r^2} \,, \\
&h(r)&= \frac{k(
    r)}{2r^2 D_f(r)}
   \l 4 \k^8 r^{13} + 2 \k^6 r^{10} r_h - 5 \k^8 r^{12} r_h - \k^6 r^9 r_h^2 \right. \nonumber \\
&& + 
     4 \k^6 r^{10} r_h \eta - 18 \k^4 r^7 r_h^2 \eta - 
     2 \k^6 r^9 r_h^2 \eta + 4 \k^2 r^4 r_h^3 \eta \nonumber \\
     && + 
     18 \k^4 r^6 r_h^3 \eta - 5 \k^2 r^3 r_h^4 \eta - 
     18 \k^4 r^7 r_h^2 \eta^2 + 12 \k^2 r^4 r_h^3 \eta^2 \nonumber \\
     && + 
     18 \k^4 r^6 r_h^3 \eta^2 + 2 r r_h^4 \eta^2 - 
     15 \k^2 r^3 r_h^4 \eta^2 - r_h^5 \eta^2  \nonumber \\
     && + 
     8 \k^2 r^4 r_h^3 \eta^3 + 4 r r_h^4 \eta^3 - 
     10 \k^2 r^3 r_h^4 \eta^3 - 2 r_h^5 \eta^3  \nonumber \\
     && \left. + 2 r r_h^4 \eta^4 - 
     r_h^5 \eta^4 \r   \,,
\eea
where 
\be
 D_f(r) \equiv (\k^2 r^3 - r_h \eta) \left[ \k^2 r^3 - r_h(1 + 
      \eta) \right] \left[ \k^4 r^6 - r_h^2 \eta(1+\eta)  \right]  ,
\ee
and
\be
k(r)=\frac{i \k^2 r^3}{(r-r_h)\sqrt{\k^4 r^6-\eta(1+\eta)r_h^2}} \,.
\ee
The most important result of using this Zerilli procedure is the value of the potential $V(r)$. We find the explicit expression 
\begin{widetext}
\be \label{Vpot}
V(r) = \frac{(r-r_h)N(r;\kappa,\eta)}{r^4 [\kappa^2 r^3 - r_h(1 + \eta)]^2 (\kappa^2 r^3 - r_h \eta)^2  [\kappa^4 r^6 -  r_h^2 \eta(1+\eta)]^2} \,,
\ee
\bea
&&N(r;\kappa,\eta) \equiv \left(6 \kappa^{16} r^{25} + \kappa^{18} r^{27} + 12 \kappa^{14} r^{22} r_h - 12 \kappa^{16} r^{24} r_h - 
  5 \kappa^{14} r^{21} r_h^2 - 3 \kappa^{12} r^{18} r_h^3 + \kappa^{10} r^{15} r_h^4  \right. \nonumber  \\
  && + 
  24 \kappa^{14} r^{22} r_h \eta  - 3 \kappa^{16} r^{24} r_h \eta- 
  99 \kappa^{12} r^{19} r_h^2 \eta  - 15 \kappa^{14} r^{21} r_h^2 \eta  + 
  30 \kappa^{10} r^{16} r_h^3 \eta  + 110 \kappa^{12} r^{18} r_h^3 \eta  - 
  15 \kappa^8 r^{13} r_h^4 \eta  \nonumber  \\
  &&  - 49 \kappa^{10} r^{15} r_h^4 \eta  + 
  21 \kappa^8 r^{12} r_h^5 \eta- \kappa^6 r^9 r_h^6 \eta  - 
  99 \kappa^{12} r^{19} r_h^2 \eta ^2 - \kappa^{14} r^{21} r_h^2 \eta^2 + 
  90 \kappa^{10} r^{16} r_h^3 \eta ^2 + 127 \kappa^{12} r^{18} r_h^3 \eta^2  \nonumber  \\
  && + 
  45 \kappa^8 r^{13} r_h^4 \eta^2 - 161 \kappa^{10} r^{15} r_h^4 \eta^2- 
  18 \kappa^6 r^{10} r_h^5 \eta^2 - 17 \kappa^8 r^{12} r_h^5 \eta^2 + 
  6 \kappa^4 r^7 r_h^6 \eta^2 + 26 \kappa^6 r^9 r_h^6 \eta^2 - 
  9 \kappa^4 r^6 r_h^7 \eta^2  \nonumber  \\
  && + 60 \kappa^{10} r^{16} r_h^3 \eta^3 + 
  13 \kappa^{12} r^{18} r_h^3 \eta^3+ 120 \kappa^8 r^{13} r_h^4 \eta^3 - 
  126 \kappa^{10} r^{15} r_h^4 \eta^3 - 72 \kappa^6 r^{10} r_h^5 \eta^3 - 
  82 \kappa^8 r^{12} r_h^5 \eta^3  \nonumber  \\
  && - 27 \kappa^4 r^7 r_h^6 \eta^3 + 
  126 \kappa^6 r^9 r_h^6 \eta^3 + 12 \kappa^2 r^4 r_h^7 \eta^3- 
  7 \kappa^4 r^6 r_h^7 \eta^3 - 9 \kappa^2 r^3 r_h^8 \eta^3 + 
  60 \kappa^8 r^{13} r_h^4 \eta^4 - 15 \kappa^{10} r^{15} r_h^4 \eta^4  \nonumber  \\
  && - 
  90 \kappa^6 r^{10} r_h^5 \eta^4 - 45 \kappa^8 r^{12} r_h^5 \eta^4 - 
  111 \kappa^4 r^7 r_h^6 \eta^4 + 183 \kappa^6 r^9 r_h^6 \eta^4 + 
  60 \kappa^2 r^4 r_h^7 \eta^4 + 16 \kappa^4 r^6 r_h^7 \eta^4 - 
  45 \kappa^2 r^3 r_h^8 \eta^4  \nonumber  \\
  && + r_h^9 \eta^4 - 
  36 \kappa^6 r^{10} r_h^5 \eta^5 - \kappa^8 r^{12} r_h^5 \eta^5 - 
  117 \kappa^4 r^7 r_h^6 \eta^5+ 97 \kappa^6 r^9 r_h^6 \eta^5 + 
  108 \kappa^2 r^4 r_h^7 \eta^5 + 8 \kappa^4 r^6 r_h^7 \eta^5 - 
  79 \kappa^2 r^3 r_h^8 \eta^5  \nonumber  \\
  &&  + 4 r_h^9 \eta^5 - 
  39 \kappa^4 r^7 r_h^6 \eta^6 + 13 \kappa^6 r^9 r_h^6 \eta^6 + 
  84 \kappa^2 r^4 r_h^7 \eta^6 - 15 \kappa^4 r^6 r_h^7 \eta^6 - 
  57 \kappa^2 r^3 r_h^8 \eta^6 + 6 r_h^9 \eta^6 \nonumber \\
  && \left. + 
  24 \kappa^2 r^4 r_h^7 \eta^7 - 9 \kappa^4 r^6 r_h^7 \eta^7 - 
  12 \kappa^2 r^3 r_h^8 \eta^7 + 4 r_h^9 \eta^7 + 
  2 \kappa^2 r^3 r_h^8 \eta^8 + r_h^9 \eta^8\right) \,.
\eea
\end{widetext}
As already announced, the coupling constant $c_{34}$ does not enter the potential $V(r)$ (nor the system \eqref{2sys} and the algebraic constraints \eqref{AC6}). 

One can rewrite this potential in the following form,
\bea \label{Vstruc}
V(r; r_h, \eta, \k)&=&\left( 1-\frac{r_h}{r} \right) \left[ \frac{r_h}{r^3}+\k^2 \right. \nonumber \\
&& + \k^2\frac{6r_h(r_h-2r) + 6r^3(r-2r_h)\k^2}{(r_h + r^3 \k^2)^2} \nonumber \\
&& \left. + (\eta+1)\k^2 V_{{\rm add}}(r; r_h, \eta, \k) \right] \,,
\eea
where $V_{{\rm add}}(r; r_h, \eta, \k)$ is a rational function of its arguments that possesses the following properties: 
(i) it has a finite limit when $r \to r_h$; (ii) it goes to zero $\sim r^{-3}$ when $r\to \infty$; (iii) it has a finite limit as $\k \to 0$ and (iv) it has a finite limit as $\eta \to -1$. 

One can then deduce a few conclusions from the rewriting \eqref{Vstruc}.
First, $V(r) \to \k^2$ when $r\to \infty$. This is related to the fact that, far away from a black hole, the wave equation describing the massive excitations of torsion bigravity satisfies a Fierz-Pauli (massive spin-2) equation \cite{Nikiforova:2009qr}. 
Second, $V(r) \to 0$ when $r\to r_h$. This is related to the fact that, very near the horizon, a massive spin-2 perturbation (having a finite frequency seen from infinity) propagates as if it was a massless spin-2 one.  This property holds also in bimetric gravity \cite{Brito:2013wya}.
 
 The third observation concerns the formal limit $\eta \to -1$. In this limit, the last term in \eqref{Vstruc} [$\propto (\eta+1)\k^2 V_{{\rm add}}$] equals to zero. The remaining terms yield the potential of the massive spin-2 field in the bimetric gravity exhibited in \cite{Brito:2013wya}, namely, 
 \bea \label{Vbcp}
 &&V_{{\rm bimetric \,gravity}}=\left( 1-\frac{r_h}{r} \right) \left[ \frac{r_h}{r^3}+\k^2 \right. \nonumber \\
&& \left. + \k^2\frac{6r_h(r_h-2r) + 6r^3(r-2r_h)\k^2}{(r_h + r^3 \k^2)^2} \right] \,.
 \eea
 This property of the formal limit $\eta\to -1$ follows from the fact exhibited in Eq.~\eqref{Wcoupling} that, in torsion bigravity, the massive spin-2 excitation has (compared to bimetric gravity) an additional coupling to the Weyl curvature proportional to $(1+\eta)$. Considering the limit $\eta \to -1$ is useful for giving checks of our results. In particular, it is easy to check that the $\eta \to -1$ limit of the matrix entries $f(r)$, $g(r)$, $h(r)$, $k(r)$ of the Zerilli transformation \eqref{ZTransform} coincides with the corresponding bimetric gravity result, as given above Eq. (30) in Ref.~\cite{Brito:2013wya}. 
 
 In addition, the first line in Eq.~\eqref{Vstruc} gives the Zerilli-like potential describing the spherically-symmetric fluctuations of a {\it massive} scalar field in a \sch background, namely,
 \be \label{VZerilliK}
 V_{\rm massive \, scalar} = \left( 1-\frac{r_h}{r} \right) \left[ \frac{r_h}{r^3}+\k^2 \right] \,.
 \ee
 Finally, the massless limit of Eq.~\eqref{Vstruc}, $\k \to 0$, namely,
 \be \label{VZerilli}
 V_{\k=0} = \left( 1-\frac{r_h}{r} \right) \frac{r_h}{r^3} \,,
 \ee
coincides with the massless limit of the scalar potential \eqref{VZerilliK}.
  The same feature holds for the bimetric gravity case \cite{Brito:2013wya} (as one can easily see in Eq.~\eqref{Vbcp}).

\section{Properties of the Zerilli-like potential $V(r)$ for torsion bigravity}
\subsection{Denominators and singularities}
The denominator of the potential for torsion bigravity, $V(r)$, reads
\be \label{denV}
r^4 [\kappa^2 r^3 - r_h(1 + \eta)]^2 (\kappa^2 r^3 - r_h \eta)^2  [\kappa^4 r^6 -  r_h^2 \eta(1+\eta)]^2 \,.
\ee
This denominator has double zeroes for three values of $r>0$. Namely, the first bracket in Eq.~\eqref{denV} 
has a double zero at 
\be  \label{r3}
r=r_3\equiv \left[ \frac{r_h(1+\eta)}{\k^2} \right]^{1/3} \,;
\ee
the second bracket has a double zero at 
\be \label{r3add}
r=r_{3{\rm add}}\equiv \left[ \frac{r_h\eta}{\k^2} \right]^{1/3} \,;
\ee
and the third bracket has a double zero at
\be \label{r6}
r=r_{6}\equiv \left[ \frac{r_h^2\eta(1+\eta)}{\k^4} \right]^{1/6} \,.
\ee
When $\eta>0$ (which is a necessary condition for the physical consistency of torsion bigravity \cite{Sezgin:1981xs}) the values of $r_3$, $r_{3{\rm add}}$ and $r_6$ are such that $r_{3{\rm add}}<r_6<r_3$. Each one of these three values (or, equivalently, the three corresponding points $r_{*3{\rm add}}<r_{*6}<r_{*3} $ on the $r_*$-axis) can potentially induce a singular behavior in the generic solution $\varphi_{\omega}(r_*)$ of Eq.~\eqref{EqPhi}. Let us study the behavior of the generic solution $\varphi_{\omega}(r_*)$ of Eq.~\eqref{EqPhi} near these three potentially singular points.

Let us start with the outermost value $r_3$. When $r\to r_3$ or, equivalently, when $r_*\to r_{*3}$, the asymptotic behavior of the potential $V[r(r_*)]$ near $r_*=r_{*3}$ is of the form
\be
V(r(r_*); r_h,\eta,\k)\simeq \frac{C^{{\rm pole}}(r_h,\eta, \k)}{(r_*-r_{*3})^2}\,.
\ee
Taking into account that $\frac{dr_*}{dr}= \frac{r}{r-r_h}$, one computes the value of the coefficient 
$C^{{\rm pole}}(r_h,\eta, \k)$ as being
\bea \label{C3}
&&C^{{\rm pole}}(r_h,\eta, \k) \nonumber \\
&&= \left[V(r; r_h,\eta,\k)(r-r_3)^2\left(\frac{r}{r-r_h}\right)^2\right]_{r=r_3}\nonumber \\
&&=2\,.
\eea
Following the usual Fuchsian analysis, one looks for asymptotic solutions of Eq.~\eqref{EqPhi} of the form 
\be \label{SolSing}
\varphi_{\omega}(r_*) \sim \left( r_*-r_{*3} \right)^s \,.
\ee
Inserting the latter power-law ansatz in Eq.~\eqref{EqPhi}, taking into account the singular behavior \eqref{C3}, one finds the following indicial equation for $s$:
\be
s(s-1) = C^{{\rm pole}}(r_h,\eta, \k) \,.
\ee
In the specific case of the $r_3$ double pole, the value $C^{{\rm pole}}(r_h,\eta, \k)=2$, Eq.~\eqref{C3}. This leads to the two exponents, $s_+=2$ and $s_\_=-1$. In other words, the generic solution $\varphi_{\omega}(r_*)$ near $r_*=r_{*3}$ has the following form
\be \label{SolR3}
\varphi_{\omega}(r_*) \underset{r_* \to r_{*3}}{=} C_{2}(r_*-r_{*3})^{2} + C_{-1}(r_*-r_{*3})^{-1} \,,
\ee
where the last term exhibits a singular behavior. To be more precise, the singular solution proportional
to  $C_{-1}$ has an expansion near $r_*=r_{*3}$ of the form
\bea
&&C_{-1}\left[ (r_*-r_{*3})^{-1} +c_1 (r_*-r_{*3}) \right. \nonumber \\
&&\left.+ c_{2L} (r_*-r_{*3})^2\log{(r_*-r_{*3})} + \ldots    \right]\,,
\eea
in which $\log{(r_*-r_{*3})}$ enters starting at the $(r_*-r_{*3})^2$-level.

At this stage, we should recall that, in view of the linear vanishing of the potential $ V(r) \propto (r-r_h)$ near the horizon, corresponding to an exponential vanishing $V[r(r_*)] \propto \exp{(r_*/r_h)}$ as $r_* \to -\infty$, the generic solution of Eq.~\eqref{EqPhi} near the horizon is of the form 
\be
\varphi_{\omega}(r_*) \underset{r_* \to - \infty}{=} C_+(\omega)e^{+ i \omega r_*} + C_\_(\omega)e^{- i \omega r_*} \,.
\ee
It is easily seen that, when completing this result by the factor $e^{-i \omega t}$, the $C_+$-term represents a wave which is outgoing from the horizon. We should therefore impose the usual no-outgoing-wave black hole boundary condition $C_+(\omega)=0$. The latter boundary condition is sufficient for determining the solution $\varphi_{\omega}(r_*)$ modulo an irrelevant overall factor (at least, in the scattering regime $\omega^2 > V(+\infty)=\k^2$). We therefore cannot impose one more boundary condition  at $r_{*3}$ to cancel the singular term  $C_{-1}(r_*-r_{*3})^{-1}$. The only way to avoid the generic presence of a singularity  at $r_{*3}$ in the torsion bigravity
master field $\varphi(r_*,t) $ is to restrict the value of the spin-2 mass $\k$ so that $r_3(r_h, \eta,\k)$  lies
 under the horizon: $r_3 < r_h$. In view of Eq. \eqref{r3}, this means constraining $\k$ to satisfy the inequality 
\be \label{CondR3}
\k >\frac{\sqrt{1+\eta}}{r_h} \,.
\ee

We have also analyzed the singular behaviors near the points $r_{*6}$ and $r_{*3{\rm add}}$. The corresponding values of the coefficient $C^{{\rm pole}}$ entering the double pole are $C^{{\rm pole}}_{6}=-\frac{1}{4}$ and $C^{{\rm pole}}_{3{\rm add}}=2$, respectively. The corresponding generic solutions are both singular. However, since we have (when $\eta>0$) the inequalities $r_{3{\rm add}}<r_6<r_3$, the satisfaction of the condition \eqref{CondR3} is enough for ensuring that all the singular points are hidden under the horizon, so that the generic solution $\varphi_{\omega}(r_*)$ will be regular outside a black hole. 

In addition, when computing (by using the inverse of Eq.~\eqref{ZTransform}) the linear perturbations in $K$ and $\tilde{H}$ in terms of $\varphi$ and its derivative, we have found that, near the singular point $r_3$, $\tilde{H}(r)$ and, therefore, $H_1(r)$, have a singular behavior of the type $\tilde{H}(r) \sim (r-r_3)^{-1}$, while $K(r)$ is finite at $r_3$ but $\frac{d K(r)}{dr}$ is logarithmically infinite.   As $H_1=F^{(1)}_{(\hat{0}\hat{1})}$ is an invariantly defined   linear perturbation of the  Ricci tensor of the torsionful connection $A$, this shows that the singularity at $r_3$ has a gauge-invariant meaning. As another way to see the gauge-invariant meaning of this singular behavior, we have checked that the (invariantly defined) torsion component ${T^{\hat{0}}}_{\hat{1}\hat{0}}$ has also singular behavior of the type ${T^{\hat{0}}}_{\hat{1}\hat{0}} \sim (r-r_3)^{-1}$, while the (invariantly defined) torsion component ${T^{\hat{1}}}_{\hat{1}\hat{0}}$ has the stronger singular behavior, ${T^{\hat{1}}}_{\hat{1}\hat{0}} \sim (r-r_3)^{-2}$. [By contrast, the metric perturbations $\phi_o(r)$ and $\Lambda_o(r)$ turn out to be both finite near $r_3$ but still to contain mildly singular contributions of the types: $\phi_o^{{\rm sing}}(r) \sim (r-r_3)^3\log{(r-r_3)}$ and $\Lambda_o^{{\rm sing}}(r) \sim (r-r_3)^2\log{(r-r_3)}$.]

We will discuss below the phenomenological consequences of the condition \eqref{CondR3}. Let us recall here that the possible necessity of imposing a constraint of the type \eqref{CondR3} was mentioned at the end of Sec. 4 of \cite{Nikiforova:2009qr}. More precisely, the St\"uckelberg treatment of \cite{Nikiforova:2009qr} showed that, in sufficiently weak Weyl-curvature backgrounds, namely, $(1+\eta) |W_{ijkl}| \ll \k^2$, the usual Fierz-Pauli mass-term in Eq.~\eqref{Wcoupling} dominates over the additional Weyl-coupling term, so that the propagating modes are not ghosts. It left open, however, the fact that there may appear ghost modes when $(1+\eta) |W_{ijkl}| \gtrsim \k^2$.  The Weyl curvature of a (\sch) black hole is of order $r_h/r^3$, and reaches its maximum value $1/r_h^2$ on the horizon. We thereby see that, indeed, the condition \eqref{CondR3} is a precise version of the no-ghost condition $(1+\eta) |W_{ijkl}| \lesssim \k^2$ discussed in \cite{Nikiforova:2009qr}. 
 
\subsection{Plots and comparison with the scalar potential and the bimetric gravity potential}

\begin{figure}[h]
\includegraphics[scale=0.6]{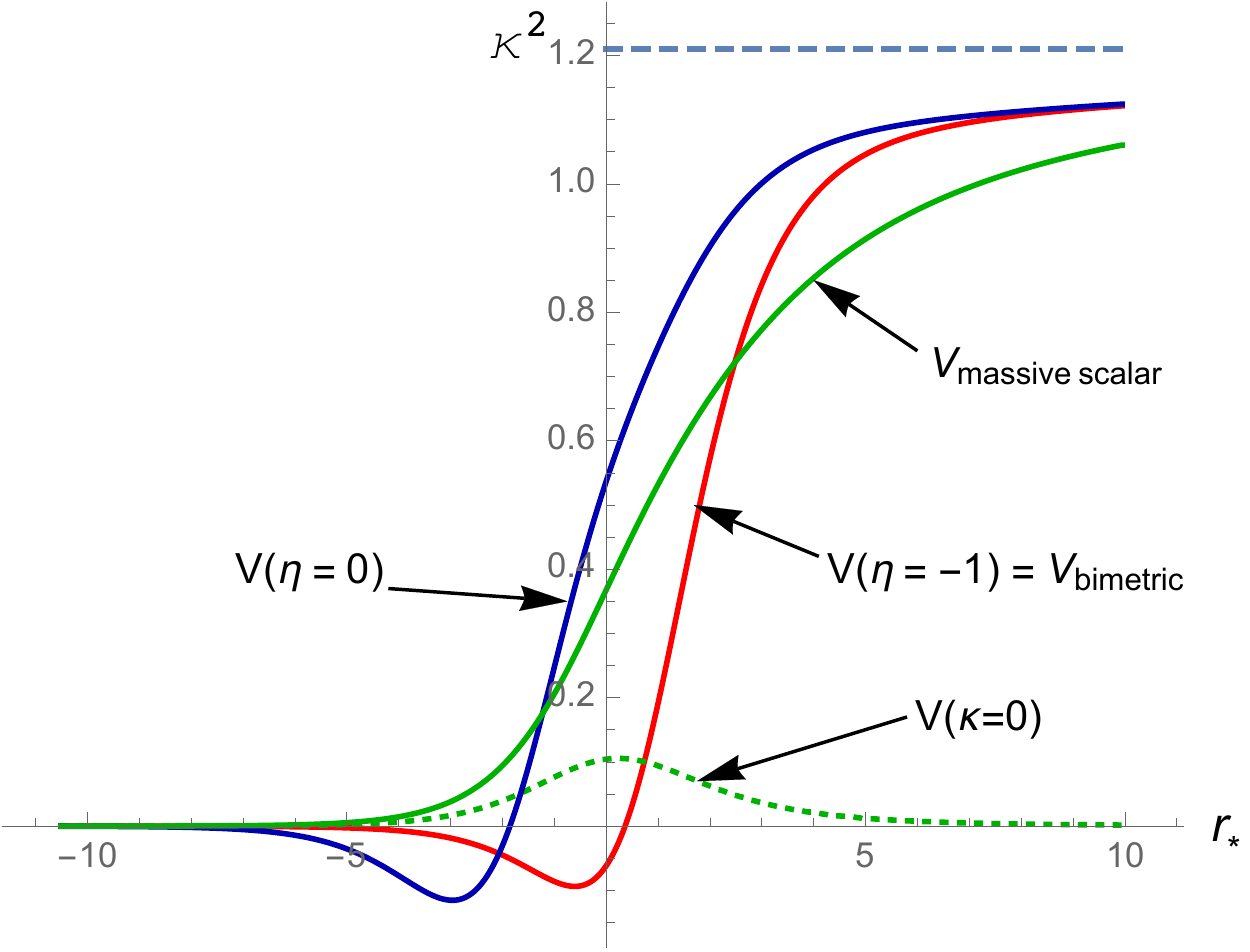}
\caption{\label{fig1}
The solid green curve $V_{{\rm massive\, scalar}}$ displays the Zerilli-like potential describing a massive scalar field on a \sch background, see Eq.~\eqref{VZerilliK}. Its massless version, Eq.~\eqref{VZerilli}, appears as the bottom dotted green curve $V(\k=0)$.
The blue curve $V(\eta=0)$ shows the potential \eqref{Vpot} of torsion bigravity computed for $\eta=0$. The red curve $V(\eta=-1)$ exhibits the potential \eqref{Vbcp} of bimetric gravity.   In all the curves, we have taken $r_h=1$ and $\k =1.1$. 
}
\end{figure}

Treating Eq.~\eqref{EqPhit} as an equation describing a (real) scalar field $\varphi(t,r_*)$, with a Lagrangian density  
\be
{\mathcal L}=\frac{1}{2}\left[\d_t\varphi(t,r_*)\right]^2 - \frac{1}{2}\left[\d_{r_*}\varphi(t,r_*)\right]^2-\frac{1}{2}V(r_*)\varphi(t,r_*)^2 \,,
\ee
one can write the conserved energy density of such a field, 
\be
{\mathcal E}=\frac{1}{2}\left[\d_t\varphi(t,r_*)\right]^2 + \frac{1}{2}\left[\d_{r_*}\varphi(t,r_*)\right]^2+\frac{1}{2}V(r_*)\varphi(t,r_*)^2 \,.
\ee
From this expression, it is clear that there are no instabilities if $V(r_*)>0$ for all $r_*$, since the energy is conserved. 
This is the case of the Zerilli-like potential describing the propagation of a scalar field on a \sch background. See Fig.~\ref{fig1} where the latter (positive) scalar potential is plotted. 

On the other hand, if the condition $V(r)>0$ is not satisfied, there might exist instabilities, at least in the case where $V(r_*)$ is {\it sufficiently} negative. 
This is the case for the even sector of the monopole perturbations in bimetric gravity. It was found in \cite{Brito:2013wya} that, for $r_h\k \leq 0.86$, there exist an instability (see a paragraph below Eq.(30) there). [Let us recall again that the potential describing the monopole perturbations in bimetric gravity is obtained from the potential \eqref{Vpot} by the formal limit $\eta \to -1$.]   Fig.~\ref{fig1} exhibits the potential \eqref{Vbcp} describing the even sector of the monopolar perturbations in bimetric gravity for $\k r_h = 1.1$. One can see that there is a region where $V_{{\rm bimetric \,gravity}}(r_*)<0$. [The negative-potential part gets deeper when $\k r_h<0.86$.]


A plot of the torsion bigravity potential for one particular set of $\eta$ and $\k$ is also exhibited in Fig.~\ref{fig1} (we chose to plot the potential for $\eta=0$, because the phenomenologically meaningful values of $\eta$ are quite small, see Sec.~\ref{phenom}). One can see that the shape of this plot is similar to that of the bimetric gravity potential. Most importantly, $V[r(r_*)]$ is not always positive, there is a region where $V[r(r_*)]<0$. The question then arises, whether the potential $V[r(r_*)]$ is sufficiently negative to create an instability or not. Below, we are going to prove that, in spite of the presence of a region of $V[r(r_*)]<0$, the potential \eqref{Vpot} provides no instabilities for solutions of the Eq.~\eqref{EqPhit}. 

\subsection{Absence of instabilities}
In the case of perturbations of a \sch black hole in bimetric gravity \cite{Brito:2013wya}, it was found: (i) that there existed instabilities for $r_h\k \lesssim 0.86$, and (ii) the complex frequency $\omega = \omega_R + i \omega_I$ of these instabilities is purely imaginary,  $\omega_R = 0$, with $\omega_I>0$. Let us prove that, if there existed instabilities in torsion bigravity, they would necessarily have also a purely imaginary frequency. 

A simple proof of this fact can be obtained by considering the conserved current of the Klein-Gordon equation \eqref{EqPhit}. Denoting $x \equiv r_*$, the latter equation can be written as
\be \label{EqPhit2}
- \frac{\d^2}{\d t^2} \varphi(t,x) + \frac{\d^2}{\d x^2} \varphi(t,x) - V[r(x)] \, \varphi(t,x)=0\,. 
\ee
For any {\it complex} solution $ \varphi(t,x)$ of Eq.~\eqref{EqPhit2} which decays both at $x \to -\infty$ and $x \to +\infty$, the following charge is conserved (if the potential $V[r(x)]$ is real)
\be \label{QDef}
Q \equiv \int_{-\infty}^{+\infty} dx \frac{i}{2} \left[ \varphi^*(t,x)\d_t \varphi(t,x) - \d_t \varphi^*(t,x)\varphi(t,x) \right] \,,
\ee
where $^*$ denotes complex conjugation. 
 
Let us suppose that there exist an unstable mode
\be \label{UnstMode}
\varphi(t,x) = e^{-i (\omega_R + i \omega_I)t }\varphi_{\omega}(x) \,.
\ee
Inserting \eqref{UnstMode} in the definition of $Q$, Eq.~\eqref{QDef}, yields
\bea
Q&=&\omega_R \int_{-\infty}^{+\infty} dx \varphi^*(t,x) \varphi(t,x) \nonumber \\
&&= \omega_R e^{2 \omega_I t}  \int_{-\infty}^{+\infty} dx |\varphi_{\omega}(x)|^2 \,.
\eea
This charge can be conserved only if the product $\omega_R \omega_I=~0$. An unstable mode ($\omega_I \neq 0$) must therefore have $\omega_R=0$. 

The search for unstable modes is thereby reduced to the search for real bound states of the Schr\"odinger-like equation 
\be
-\frac{\d^2}{\d x^2} \varphi_{\omega}(x) + V[r(x)]\varphi_{\omega}(x)=\omega^2 \varphi_{\omega}(x) \,,
\ee
where $\omega^2 = -\omega_I^2$ is {\it negative}.

In order to establish\footnote{Note that the additional potential contribution $(\eta+1)\k^2 V_{\rm add}(r)$, compared to the bimetric gravity one, is not always positive. So that we cannot establish the absence of unstable modes simply on the ground of the inequality \eqref{CondR3}.} the presence or absence of unstable modes, we can then use the theorem \cite{Hilbert, Chadan:2002gj}
saying that the number of negative-energy bound states of a potential $V(x)$ is equal to the number of nodes of the zero-energy wave function $\varphi_0(x)$ satisfying one of the bound-state boundary conditions (see Appendix \ref{Theorem} for the sketch of a proof of this theorem). In the case of torsion bigravity, where the potential tends to $+\k^2$ as $x \equiv r_* \to +\infty$, it is convenient to impose on the zero-energy wave function $\varphi_0(x)$, satisfying 
\be
-\frac{\d^2}{\d x^2} \varphi_0(x) + V[r(x)]\varphi_0(x)=0 \,,
\ee
the boundary condition that it vanishes at $+\infty$,
\be
 \varphi_0(x) \underset{x\to + \infty}{\approx} e^{-\k x} \,.
\ee
A numerical study of the so-defined wave-function $\varphi_0(x)$ for all relevant values of $\eta$ and $\k$, i.e., $\eta>0$ and $(\k r_h)^2>1+\eta$, has shown that this wave function stays positive for all values of $x$. In other words, the number of nodes is zero, which proves that there are no negative-energy bound states.

By contrast, we have checked the usefulness of this theorem by applying the same method to the potential $V_{\rm bimetric \, gravity}$, \eqref{Vbcp}. We indeed confirmed that, when $\k r_h \leq 0.86$, the right decaying zero-energy wave function does have a single node, thereby proving the existence of one bound state, i.e., one unstable mode.

\section{Phenomenological consequences of our results}\label{phenom}

We found that torsion bigravity perturbations of \sch black holes were developing singularities if $\k^2 < (1+\eta)/r_h^2$. More generally, we have seen above that singularities might develop when $\k^2$ is smaller than $(1+\eta)$ times the eigenvalues of the Weyl curvature. The astrophysical objects having the largest Weyl curvature would be small-mass black holes. In absence of experimental evidence for the existence of sub-solar-mass black holes we shall conservatively assume that the largest Weyl curvature\footnote{Though early stages of cosmological expansion feature large curvatures, these are not large Weyl curvatures because Friedmann models are conformally flat.} to consider is the one at the surface of a $2M_\odot$ black hole. [Indeed, there are no known neutron stars for which the Weyl curvature would be larger than the Weyl curvature at the surface of a $2M_\odot$ black hole.] This yields the phenomenological constraint
\be \label{ConstrV}
\k > \frac{\sqrt{1+\eta}}{6} \;{\rm km}^{-1} \,.
\ee
Remembering that $\eta$ must be positive, this means that the range $\k^{-1}$ of the massive spin-2 excitation must be smaller than 6 km. If we assume that the range is indeed of order of a few km, the existing gravitational tests then imply (see Section X.A in \cite{Damour:2019oru}) that 
\be
\eta \lesssim 3\times 10^{-4} \quad \text{for} \quad \k^{-1} \lesssim 10\, \text{km}\,. 
\ee
The schematic structure of the torsion bigravity action \eqref{lag0} reads
\be 
L \sim c_R R+  c_F F\l 1+ \frac{1+\eta}{\k^2}F \r \,,
\ee
where the (schematic) term $\frac{1+\eta}{\k^2}F$ is a higher derivative fractional correction to the 
Einstein-Cartan-like $F$-term. The development of singularities when $\k^2 \lesssim (1+\eta)|W|$ then appears as being associated to situations where the fractional correction $\frac{1+\eta}{\k^2}F$ becomes of order unity. From the theoretical point of view, we might then expect that this signals the necessity of completing the torsion bigravity action by higher-order-in-curvature terms, say, 
  \be \label{lagUV}
L_{UV} \sim c_R R+  c_F F \hspace{-0.7mm} \l \hspace{-0.4mm} 1+ \frac{1+\eta}{\k^2}F + \left[\frac{F}{\k^2}\right]^2\hspace{-0.9mm}+\left[\frac{F}{\k^2}\right]^3 \hspace{-0.9mm}+ ... \r \hspace{-0.5mm}.
\ee
Such an extended model (possibly of the Born-Infeld type \cite{Born:1934gh} or another limiting-curvature model \cite{Mukhanov:1991zn}) might cure the mass-related singularity while, hopefully, improving the UV-behavior of the theory.

\section{Conclusions}
We studied spherically-symmetric perturbations of \sch black holes within torsion bigravity theories. These Einstein-Cartan-type theories (with dynamical torsion) contain two excitations:
an Einsteinlike massless spin-2 one, and a massive spin-2 one, of inverse range $\k$. 

We proved that the odd-parity time-dependent spherically-symmetric perturbed sector is trivial (see Appendix \ref{odd}). 
We reduced the full set of perturbed even-parity equations to a system of two first-order differential equations (see Eqs.~\eqref{2sys}), together with six algebraic constraints \eqref{AC6}. [This confirms the absence of a Boulware-Deser sixth degree of freedom.] 

We then showed how to transform the system \eqref{2sys} of two first-order differential equations into a Zerilli-like equation 
\be
 \frac{\d^2}{\d r_*^2} \varphi(t,r_*) - \frac{\d^2}{\d t^2} \varphi(t,r_*)= V[r(r_*)] \, \varphi(t,r_*)\,
\ee
with potential $V(r)$ given in Eq.~\eqref{Vpot}. 
Several features of this potential were discussed. In particular, it was shown that it reduces to the corresponding potential \eqref{Vbcp} in ghost-free bimetric gravity \cite{Brito:2013wya} when the torsion bigravity coupling constant $\eta \equiv c_F/c_R$  formally takes the value $\eta=-1$. [This is related to the fact that, in torsion bigravity, the Fierz-Pauli-like equation describing massive spin-2 excitations are modified by an extra coupling to the Weyl curvature, proportional to $1+\eta$ (see Eq.~\eqref{Wcoupling}).]

On the other hand, contrary to the bimetric gravity potential \eqref{Vbcp}, the torsion bigravity potential, considered for physically allowed values $\eta>0$, contains possibly vanishing denominators outside the horizon when $(\k r_h)^2 < 1+\eta$. It was shown that these denominators, if present, would induce a corresponding singular behavior in the generic solution of the Zerilli-like equation. It was then concluded that a necessary condition for the physical acceptability of torsion bigravity is to constrain the mass of the spin-2 excitation by the condition 
\be \label{ConstrK}
\k^2 > \frac{1+\eta}{r_h^2} \,,
\ee
where $r_h$ denotes the radius of the considered black hole. [The condition \eqref{ConstrK} is again linked to the presence of an additional Weyl-curvature coupling in  the Fierz-Pauli-like equation, see Eq.~\eqref{Wcoupling},  describing massive spin-2 excitations in torsion bigravity.]

The torsion bigravity potential $V(r; \eta, \k)$, now considered for physically allowed values $\eta>0$ and $(\k r_h)^2 > 1+\eta$, is not everywhere positive (see Fig.\ref{fig1}). We could, however, prove the stability of \sch black holes against monopolar perturbations, by (numerically) showing the absence of negative-energy bound states in the potential $V(r; \eta, \k)$ (using the theorem sketched in Appendix \ref{Theorem}). 

The constraint \eqref{ConstrK} has important consequences for phenomenological applications of torsion bigravity. The first consequence is that one cannot consider large inverse ranges $\k^{-1}$, say, of galactic or cosmological sizes. The constraint \eqref{ConstrK} restricts the physical applicability of torsion bigravity to inverse ranges smaller or equal to $\sim 6 \;{\rm km}$ (see \eqref{ConstrV}).  Even when taking into account such a constraint, torsion bigravity could still have important phenomenological consequences for the physics of neutron stars and (stellar-mass) black holes. We leave a discussion of these phenomenological consequences to future work.

\section*{Acknowledgments}
The author thanks Thibault Damour, Henri Epstein and Valery Rubakov for useful suggestions.

\appendix

\section{Field equations in general form, even sector} \label{apA}
The field equations for general $R+R^2$ Riemann-Cartan-type theories (with generic metric compatible affine connection), in the time-dependent spherically-symmetric case, were given by Rauch and Nieh in \cite{Rauch:1981tva}. The original field equations that describe the even-parity sector are the ``gravity'' equations (4.3a), (4.3b), (4.3c), (4.3d), (4.3e), and the ``connection'' equations (4.5a), (4.5b), (4.5c), (4.5d) of  \cite{Rauch:1981tva}.  [By contrast to these original field equations, beware of some sign misprints in the ``slightly different form'' given later for the field equations, namely, Eqs.~(4.4a)--(4.4f) and, consequently, Eqs.~(6.2a)--(6.2e). Specifically, the signs of the contributions proportional to $\frac{1}{3\lam}(-a+2c-3\lam)$ in Eqs.~(4.4c) and (6.2c) should be reversed.]  To adjust these field equations to the case of torsion bigravity, one needs to take the following values of the parameters $\lam_{RN}, a_{RN}, b_{RN}, c_{RN}, p_{RN}, q_{RN}, r_{RN}, s_{RN}, t_{RN}$ used in \cite{Rauch:1981tva} (see Eq.~(1.1) there for definitions of these parameters):

\bea \label{RNParam}
&& \lam_{RN}  = c_F +c_R \,,\nonumber \\
&&c_{RN}=b_{RN}=-a_{RN}=c_F \,, \nonumber \\
&& p_{RN}= q_{RN}=0 \,, \nonumber \\
&& s_{RN}=\frac{r_{RN}}{2}=-\frac{c_{F^2}}{6} \,, \nonumber \\
&& t_{RN}= \frac{c_{34}}{2} + \frac{2 c_{F^2}}{3} \,.
\eea

Let us recall the field variables in even-parity sector: $\Phi(t,r)$ and $\Lambda(t,r)$ describing the spherically symmetric metrics, and four variables describing dynamical torsion, 
\bea \label{defTorVar}
V(t,r) & \equiv &{A^{\hat{r}}}_{\hat{t}\hat{t}}=+{A^{\hat{t}}}_{\hat{r}\hat{t}} = e^{-\Lambda}\Phi^{\prime} + {T^{\hat{t}}}_{\hat{r}\hat{t}}  \,,  \nonumber \\
W(t,r) & \equiv &{A^{\hat{r}}}_{\hat{\theta}\hat{\theta}}={A^{\hat{r}}}_{ \hat{\phi}\hat{\phi}}=-{A^{\hat{\theta}}}_{\hat{r}\hat{\theta}} = -{A^{\hat{\phi}}}_{\hat{r}\hat{\phi}} \nonumber \\
&& = - \frac{e^{-\Lambda}}{r} - {T^{\hat{\theta}}}_{\hat{r}\hat{\theta}}  \,,  \nonumber \\
X(t,r) & \equiv &{A^{\hat{r}}}_{\hat{t}\hat{r}}=+{A^{\hat{t}}}_{\hat{r}\hat{r}} = e^{-\Phi}\d_t \Lambda + {T^{\hat{r}}}_{\hat{t}\hat{r}}   \,,  \nonumber \\
Y(t,r) & \equiv &{A^{\hat{\theta}}}_{\hat{t}\hat{\theta}}=+{A^{\hat{t}}}_{\hat{\theta}\hat{\theta}} ={A^{\hat{\phi}}}_{\hat{t}\hat{\phi}}=+{A^{\hat{t}}}_{\hat{\phi}\hat{\phi}}   \nonumber \\
&& = {T^{\hat{\theta}}}_{\hat{t}\hat{\theta}}   \,.  \nonumber \\
\eea
Introducing the following auxiliary quantities,
\bea
&&{\bf A} \equiv -e^{-\Lambda-\phi}\left[\d_t(e^{\Lambda}X)  - \d_r(e^{\phi}V)\right] \,, \nonumber \\
 && {\bf C} \equiv -\d_t Y e^{-\Phi} - V W \,, \nonumber  \\
 && {\bf D} \equiv \frac{e^{-\Lambda}}{r} \l r Y \r^{\prime} + X W \,, \nonumber  \\
 && {\bf G} \equiv -\d_t W e^{-\Phi} - V Y \,, \nonumber  \\
 && {\bf H} \equiv \frac{e^{-\Lambda}}{r} \l r W \r^{\prime} + X Y \,, \nonumber  \\
 && {\bf L} \equiv \frac{1}{r^2} + Y^2 - W^2 \,, \nonumber  \\
 && {\bf \Omega} \equiv {\bf A} - {\bf L} + 2({\bf C}-{\bf H}) \,,
\eea
and using \eqref{RNParam} and \eqref{defTorVar}, one can rewrite the 9 (nonlinear) field equations (4.3a), (4.3b), (4.3c), (4.3d), (4.3e), (4.5a), (4.5b), (4.5c) and (4.5d) of \cite{Rauch:1981tva}, adapted for torsion bigravity, in the following form:
\begin{widetext}
\bea \label{EG1LinOsc1}
&& \frac{c_{34} + c_{F^2}}{c_F + c_R} ( {\bf G}^2 -{\bf D}^2 ) + 2 {\bf H} + 
 {\bf L}  - \frac{c_{F^2}}{3 (c_F + c_R)} \left[({\bf A} -  
   {\bf C})^2 - ({\bf H} - {\bf L})^2\right] + \frac{2 c_R}{c_F + c_R} \left[\textcolor{white}{\frac{i}{i}}{T^{\hr}}_ {\htt\hr} {T^{\htheta}}_ {\htt\htheta} \right. \nonumber \\
&& \left. + 
   \frac12 \left({{T^{\htheta}}_ {\hr\htheta}}^2 + {{T^{\htheta}}_ {\htt\htheta}}^2\right)  - 
   {T^{\htheta}}_ {\htt\htheta} X - \left({T^{\hr}}_ {\htt\hr} + {T^{\htheta}}_ {\htt\htheta}\right) Y + e^{-\Lambda}\left(2r^{-1}{T^{\htheta}}_ {\hr\htheta} +\d_r {T^{\htheta}}_ {\hr\htheta}\right) \textcolor{white}{\frac{i}{i}}\right]  + {\rm odd}^2=0 \,,
\eea

\bea \label{EG2LinOsc1}
&&\frac{c_{34} + c_{F^2}}{c_F + c_R} ({\bf G}^2 - {\bf D}^2) + 2 {\bf C} - 
 {\bf L} + 
 \frac{c_{F^2}}{3 (c_F + c_R)} [({\bf A} + {\bf H})^2 - ({\bf C} + 
       {\bf L})^2] + \frac{2 c_R}{
  c_F + c_R} \left[e^{-\Phi} \d_t {T^{\htheta}}_{\htt\htheta}   \right. \nonumber \\
  && \left. + 
    \frac12 \l {T^{\htheta}}_{\hr\htheta}^2 + {T^{\htheta}}_{\htt\htheta}^2 \r + {T^{\htheta}}_{\hr\htheta} {T^{\htt}}_{\hr\htt} - 
    {T^{\htheta}}_{\hr\htheta} V + \l {T^{\htheta}}_{\hr\htheta} + {T^{\htt}}_{\hr\htt} \r W \right] + {\rm odd}^2 = 0 \,,
\eea

\bea \label{EG3LinOsc1}
&&{\bf A} + {\bf C} - {\bf H} - 
 \frac{c_{F^2}}{ 3 (c_F + c_R)} ({\bf A} - {\bf C} + {\bf H} - 
    {\bf L}) ({\bf A} + {\bf L}) + 
 \frac{c_R}{c_F + c_R}  \left\{{T^{\hr}}_{\htt\hr} {T^{\htheta}}_{\htt\htheta} - {T^{\htheta}}_{\hr\htheta} {T^{\htt}}_{\hr\htt} \right. \nonumber \\
 && + 
    \frac{1}{r} e^{-\Lambda - \Phi} \left[ \left(\d_t {T^{\hr}}_{\htt\hr} + \d_t {T^{\htheta}}_{\htt\htheta}\right) e^\Lambda r - \left(\d_r {T^{\htheta}}_{\hr\htheta} + \d_r {T^{\htt}}_{\hr\htt}\right) e^\Phi  r + 
      e^\Lambda r \d_t\Lambda \left({T^{\hr}}_{\htt\hr}+ {T^{\htheta}}_{\htt\htheta} \right) - 
       e^\Phi \left({T^{\htheta}}_{\hr\htheta}  + {T^{\htt}}_{\hr\htt} \right) \right. \nonumber \\
       && \left. \left.- 
      e^\Phi r  \d_r \Phi \left({T^{\htheta}}_{\hr\htheta} + {T^{\htt}}_{\hr\htt} \right)\right]\right\} + {\rm odd}^2 = 0 \,,
\eea

\bea \label{EG4LinOsc1}
&& {\bf D} - \frac{c_{34} + c_{F^2}}{c_F + c_R} ({\bf C}{\bf D}- 
    {\bf G}{\bf H}) + 
 \frac{c_{34}}{c_F + c_R} {\bf D}({\bf C}- {\bf H}) + 
 \frac{c_{F^2}}{3 (c_F + c_R)} {\bf D}{\bf \Omega}    \nonumber \\
    && - 
 \frac{c_R}{c_F + c_R} \left[ 
    \frac{e^{-\Lambda}}{r} \l r\d_r{T^{\htheta}}_{\htt\htheta} + 2 {T^{\htheta}}_{\htt\htheta} \r  + \left({T^{\hr}}_{\htt\hr}  + {T^{\htheta}}_{\htt\htheta} \right) \left({T^{\htheta}}_{\hr\htheta} + W \right) - 
    {T^{\htheta}}_{\hr\htheta} X \right] + {\rm odd}^2 = 0 \,,
\eea

\bea \label{EG5LinOsc1}
&& {\bf G}- \frac{c_{34} + c_{F^2}}{c_F + c_R} ({\bf C}{\bf D}- 
    {\bf G}{\bf H}) + 
\frac{c_{34}}{c_F + c_R} {\bf G} ({\bf C}- {\bf H}) + 
 \frac{c_{F^2}}{3 (c_F + c_R)} {\bf G}{\bf \Omega} \nonumber \\
 && - 
 \frac{c_R}{c_F + c_R} \left[ e^{-\Phi} \d_t {T^{\htheta}}_{\hr\htheta}  - 
    {T^{\htheta}}_{\htt\htheta}  V - \left({T^{\htheta}}_{\hr\htheta}  + {T^{\htt}}_{\hr\htt} \right) \left(-{T^{\htheta}}_{\htt\htheta}  + Y \right) \right] + {\rm odd}^2 =0 \,,
\eea

\bea \label{ET1LinOsc1}
&&-\frac12 c_{F^2} e^{-\Lambda} \d_r{\bf \Omega}  + 
 {T^{\htheta}}_{\hr\htheta}  (3 c_F + c_{F^2} {\bf \Omega}) + 
 3 c_{34} ({\bf D} - {\bf G}) Y + 
 \frac{3}{2} c_{F^2} \left( \frac{e^{-\Lambda}}{r} (r \d_r {\bf A}  + 2 {\bf A}) + 
    2 {\bf C} W - 2 {\bf G} Y\right) \nonumber \\
    && + {\rm odd}^2 =0 \,, 
\eea

\bea  \label{ET2LinOsc1}
-\frac12 c_{F^2} e^{-\Phi} \d_t{\bf \Omega}  + 
 {T^{\htheta}}_{\htt\htheta}  (3 c_F + c_{F^2} {\bf \Omega}) - 
 3 c_{34} ({\bf D} - {\bf G}) W + 
 \frac{3}{2} c_{F^2} \left(e^{-\Phi} \d_t{\bf A}  - 2 {\bf D} W + 
    2 {\bf H} Y \right) + {\rm odd}^2 =0 \,,
\eea

\bea  \label{ET3LinOsc1}
&&3 c_F ({T^{\hr}}_{\htt\hr}  + {T^{\htheta}}_{\htt\htheta} ) - 
 c_{F^2} \left[e^{-\Phi} \d_t{\bf \Omega}  - ({T^{\hr}}_{\htt\hr}  + {T^{\htheta}}_{\htt\htheta} ) {\bf \Omega} \right] + 3 c_{F^2} \left[ e^{-\Lambda} \d_r {\bf D}  + 
    e^{-\Phi^{\textcolor{white}{\prime}}}\hspace{-0.8mm} \d_t {\bf C}  +  e^{-\Phi} {\bf C} \d_t\Lambda  \right.  \nonumber \\
    &&  \left. + 
    \frac{e^{-\Lambda}}{r} (1 + r \d_r\Phi ) {\bf D} + 
    {\bf G} V + {\bf H} X + {\bf L} Y \right] + 
 3 c_{34} \frac{e^{-\Lambda}}{r} \left[ (\d_r{\bf D} - \d_r {\bf G}) r + ({\bf D} - {\bf G}) (1 + \d_r\Phi r - 
       e^{\Lambda^{\textcolor{white}{\prime}}}\hspace{-0.8mm} r V ) \right] \nonumber \\
       && + {\rm odd}^2 = 0 \,, 
\eea

\bea \label{ET4LinOsc1}
&& 3 c_F ({T^{\htheta}}_{\hr\htheta}  + {T^{\htt}}_{\hr\htt} ) - 
 c_{F^2} \left[ e^{-\Lambda} \d_r {\bf \Omega}  - ({T^{\htheta}}_{\hr\htheta}  + 
       {T^{\htt}}_{\hr\htt} ) {\bf \Omega} \right] + 
 3 c_{34} e^{-\Phi} \left[ ({\bf D} - {\bf G}) (\d_t\Lambda - e^\Phi  X) + 
    \d_t{\bf D} - \d_t {\bf G} \right] \nonumber \\
    && + 
 c_{F^2} \left[ -3  e^{-\Phi}  {\bf G} \d_t\Lambda - 
    3 \frac{e^{-\Lambda}}{r} (1 + r \d_r\Phi) {\bf H} - 
    3 {\bf C} V - 3 {\bf L} W - 3 {\bf D} X - 
    3 e^{-\Phi} \d_t {\bf G} - 3 e^{-\Lambda} \d_r{\bf H} \right] + {\rm odd}^2 =0 \,. \nonumber \\
\eea

\end{widetext}
In Eqs.~\eqref{EG1LinOsc1}-\eqref{ET4LinOsc1}, the notation ${\rm odd}^2$ denotes terms that are quadratic in odd-parity variables, so that they do not affect the linearized equations, and, thus, have no importance for the current study.

\section{Parity-odd sector} \label{odd}
The five exact field equations that involve the odd-parity variables $C_1, \ldots, C_4$ linearly are displayed in Eqs. (4.3f), (4.6a)--(4.6d) of \cite{Rauch:1981tva} (let us recall that the parameters $a, b, c, p, q, {\rm etc}$ of \cite{Rauch:1981tva} take, in the case of torsion bigravity, the values given in Eq.~\eqref{RNParam}). Recalling that the background values $C_{i\,S}(r)$ equal to zero, we insert in the latter odd-parity equations the perturbed values
\be
C_i(t,r)=\varepsilon \hat{C}_i(t,r) \,, \quad i=1,\ldots,4 \,,
\ee
which correspond to the odd-parity sector of Eq.~\eqref{linearization}.
One so obtains five linear equations involving only the four odd-parity field variables $\hat{C}_1, \ldots, \hat{C}_4$. In what follows, we omit the notation $\;\widehat{}\;$, for simplicity. 

These linearized equations are as follows\footnote{These are linearized versions of the equations of \cite{Rauch:1981tva} written in the following order: (4.3f), (4.6a), (4.6b), (4.6c), (4.6d).}:

\bea \label{EGodd1}
&&-\frac{1}{2 r^{9/2} \sqrt{
  r - r_h} (1 + \eta) \lambda}  (c_{34} r_h + c_{34} r_h \eta \nonumber \\
  && + 
    r^3 \eta \lambda) (2 r^2 \d_r C_4  + 2 r^2 \d_t C_3  - 2 r r_h \d_r C_4  - 
    2 r C_2 \nonumber \\
    &&   + 2 r_h C_2 + 2 r C_4 - r_h C_4) = 0 \,,
\eea

\be \label{ETodd1}
\frac{3 \eta \lambda C_3}{1 + \eta} = 0 \,,
\ee

\bea \label{ETodd2}
&&\frac{3 c_{34} \d_t C_3}{r}  + \frac{3 c_{34} (r - r_h)}{r^3} (r \d_r C_4  - C_2)  \nonumber \\
&& + 
 \frac{3}{2 r^3 (1 + \eta)} \l 2 c_{34} r   - c_{34} r_h + 2 c_{34} r \right. \eta \nonumber \\
&& \left.  - c_{34} r_h \eta + 
    2 r^3 \eta \lambda \r C_4 = 0 \,,
\eea

\bea \label{ETodd3}
&&\frac{3 c_{34}}{r} \l -\d_t C_2 + r \d^2_{rt}C_4 \r \nonumber \\
&& + 
 \frac{3 c_{34}}{2 r (r - r_h)} \left(2 r \d_t C_4 + 2 r^2 \d^2_t C_3  - r_h \d_t C_4  \right) \nonumber \\
 && + 
 \frac{3 \eta \lambda}{1 + \eta} (C_1 - C_3) = 0 \,,
\eea

\bea \label{ETodd4}
&&\frac{3 c_{34}}{r^2} (-\d_r C_2 + r \d^2_r C_4) (r - r_h) \nonumber \\
&&  + 
 \frac{3 c_{34}}{r^2} (2 r \d_r C_4 + r^2 \d^2_{rt}C_3  - r_h \d_r C_4 ) \nonumber \\
 &&  + \frac{3}{4 r^3 (r - r_h) (1 + \eta)}
  \left[2 c_{34}  r^2 (2 r - 3 r_h) (1 + \eta)\d_t C_3 \right. \nonumber \\ 
  && - 
    2 (r - r_h) \left(c_{34} r_h (1 + \eta) - 2 r^3 \eta \lambda \right) C_2 \nonumber \\
    && \left. - \left(c_{34} r_h^2 (1 + \eta) + 
       4 r^3 (r_h-r) \eta \lambda \right) C_4 \right] = 0 \,.
\eea

Eq.~\eqref{ETodd1} implies  
\be \label{C3=0}
C_3(t,r)=0 \,.
\ee
Then, inserting this result into the following combination of equations, $\frac{c_{34} r_h (1 + \eta) + r^3 \eta \lambda}{r^{
 3/2} (1 + \eta) \lambda} \eqref{ETodd2} + 3 c_{34} \sqrt{r - r_h}\eqref{EGodd1}$, gives
 \be 
 \frac{3 \eta (c_{34} r_h (1 + \eta) + r^3 \eta \lambda) C_4}{r^{3/2} (1 + \eta)^2} = 0 \,,
 \ee
 which implies  
 \be \label{C4=0}
 C_4(t,r)=0 \,.
 \ee
Substituting \eqref{C3=0} and \eqref{C4=0} in Eq.~\eqref{ETodd2} then gives $C_2(t,r)=0$. Then, substituting $C_2(t,r)=0$ together with  \eqref{C4=0} and \eqref{C3=0} in Eq.~\eqref{ETodd3} gives $C_1(t,r)=0$. Finally, we get 
\be
C_1(t,r) = C_2(t,r) = C_3(t,r) = C_4(t,r) =0 \,
\ee
which also identically satisfies Eq.~\eqref{ETodd4}. 
Therefore, there are no spherically-symmetric odd-parity perturbations.

\section{Sketch of a proof of a bound-states counting theorem} \label{Theorem}
One can formulate a concrete theorem within a Sturm-Liouville context as follows:

{\it
The number of negative-energy bound states of the potential $V(x)$ between $x_0$ and $x_1>x_0$, i.e., the number of solutions with $\lam<0$ of the following problem
\be \label{eq1}
\d^2_x \varphi_{\lam}(x) = \l V(x)-\lam \r \varphi_{\lam}(x) \,, \; \varphi_{\lam}(x_0)=0 \,, \; \varphi_{\lam}(x_1)=0
\ee
is equal to the number of nodes of a (non-zero) solution of the following problem
\be \label{eq2}
\d^2_x \varphi_0(x)  = V(x) \varphi_0(x) \,, \quad \varphi_0(x_0)=0 \,.
\ee
}

The proof of this theorem is obtained by following, as $\lam$ continuously increases from a sufficiently negative value to zero, the nodes of a solution satisfying only one of the boundary conditions, say, $\varphi_{\lam}(x_0)=0$, which can be completed (without loss of generality) by the condition $\varphi^{\prime}_{\lam}(x_0)=1$.

\begin{figure}
\includegraphics[scale=0.6]{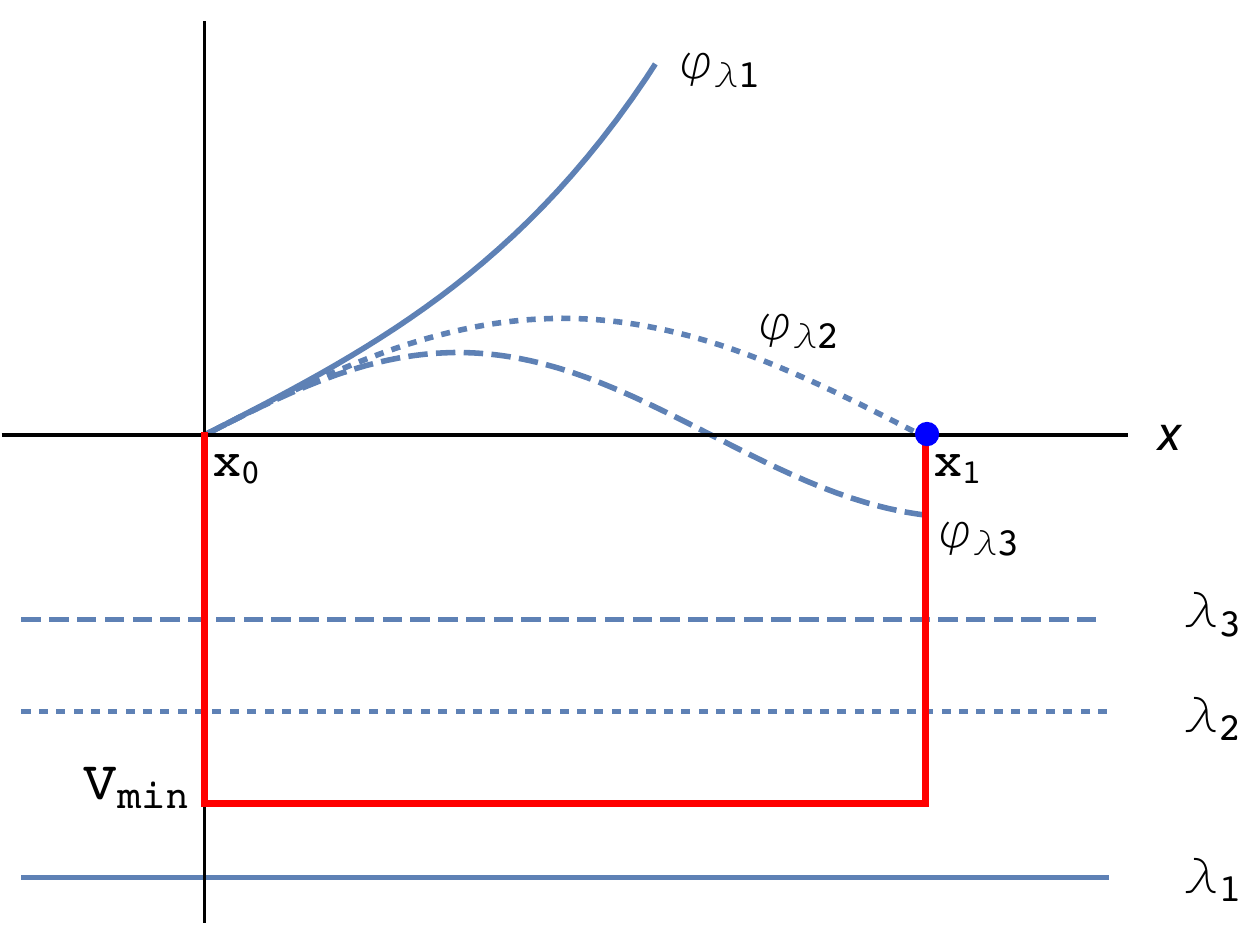}
\caption{\label{fig3}
A negative potential (in red) entering the equation $\d^2_x \varphi_{\lam}(x) = \l V(x)-\lam \r \varphi_{\lam}(x)$. The functions $\varphi_{\lam1}$, $\varphi_{\lam2}$ and $\varphi_{\lam3}$ are the three solutions corresponding to $\lam=\lam_1$, $\lam=\lam_2$ and $\lam=\lam_3$, satisfying the left boundary condition $\varphi_{\lam i}(x_0)=0$, $\varphi_{\lam i}^{\prime}(x_0)=0$. The energy parameter $\lam$ continuously increases, $\lam_1<\lam_2<\lam_3$, and the value $\lam_2$ corresponds to the first bound state of the potential. 
}
\end{figure}

First, if $\lam$ is less than the minimal value $V_{min}$ of the potential $V(x)$ (see Fig.~\ref{fig3}), the difference $V(x)-\lam$ is always positive, the second derivative $\varphi^{\prime\prime}_{\lam}(x_0)$ is positive in the right vicinity of $x_0$ (here and in what follows, the prime denotes $\d_x$). It is then easily seen that the second derivative $\varphi^{\prime\prime}_{\lam}(x_0)$ will always stay positive, so that the curve $\varphi_{\lam}(x)$ will be a convex, monotonically increasing function which will never cross zero (see upper curve, $\varphi_{\lam1}$, in Fig.~\ref{fig3}).

If now one increases $\lam$ so that $\lam>V_{min}$, the second derivative in \eqref{eq1} becomes negative at some stage. The curve $\varphi_{\lam}(x)$ becomes concave which might allow it (if $V(x)-\lam$ is sufficiently negative) to turn over and cross zero. Let us denote as $\xi_{\lam}$ the value of $x$ where $\varphi_{\lam}(x)=\varphi_{\lam}(\xi_\lam)=0$. Let us then prove that as $\lam$ is increased to a nearby value $\mu \equiv \lam + \eps$, where $\eps>0$ and small, the corresponding value $\xi_\mu$ where $\varphi_{\lam}(\xi_\mu)=0$ is always {\it on the left} of $\xi_\lam$. 

Integrating the easily checked identity 
\bea
&&\l\varphi_\mu \varphi_\lam^{\prime} - \varphi_\mu^{\prime} \varphi_\lam \r^{\prime} = \varphi_\lam \varphi_\mu\left[ V(x)-\lam-V(x)+\mu \right] \nonumber \\
&&=(\mu-\lam)\varphi_\lam(x) \varphi_\mu (x)
\eea
between $x_0$ and $\xi_\lambda$ yields 
\be
\varphi_\mu(\xi_\lam)\varphi^{\prime}_\lam(\xi_\lam) = (\mu-\lam)\int_{x_0}^{\xi_\lam} \varphi_\lam(x) \varphi_\mu (x) \,.
\ee
The latter identity is easily seen to imply that, whatever be the sign of the slope $\varphi^{\prime}_\lam(\xi_\lam)$, $\varphi_\mu(x)$ (when $\mu$ is slightly larger than $\lam$) will have a zero located at a position $\xi_\mu$ {\it on the left} of $\xi_\lam$. 

The rest of the proof consists in following this migration towards the left of the zeros of $\varphi_\lam(x)$ as $\lam$ increases. Each bound state corresponds to the case where one such zero passes through $x_1$ (as illustrated by the middle curve, $\varphi_{\lam2}$, in Fig.\ref{fig3}). As a consequence, when $\lam$ reaches zero, the zero-energy wave function $\varphi_0(x)$ has accumulated $N$ nodes in the open interval $(x_0; x_1)$, where $N$ is exactly the number of bound states of negative energy, $\lam<0$.


\end{document}